\pdfoutput=1

\PassOptionsToPackage{hyphens}{url}

\documentclass[acmsmall, screen=true, nonacm]{acmart}
\def\BibTeX{{\rm B\kern-.05em{\sc i\kern-.025em b}\kern-.08emT\kern-.1667em\lower.7ex\hbox{E}\kern-.125emX}}
\acmJournal{TOPS}
\usepackage{hyperref}
\hypersetup{%
	colorlinks=true,
	linkbordercolor=red,
	pdfborderstyle={/S/U/W 1}%
}

\usepackage{pifont}
\newcommand{\cmark}{\ding{51}}
\newcommand{\xmark}{\ding{55}}

\usepackage{framed}

\usepackage{longtable}
\usepackage{booktabs}
\usepackage{tabularx}

\newcommand*\circled[1]{\raisebox{.5pt}{\textcircled{\raisebox{-.9pt} {#1}}}}

\copyrightyear{2023}


\begin{document}

\title{The Android Platform Security Model (2023)}
\titlenote{Last updated in December 2023 based on Android 14 as released. A previous version was published in \href{https://doi.org/10.1145/3448609}{ACM Transactions on Privacy and Security} from April 2021.}

\author{Ren\'e Mayrhofer}
\orcid{0000-0003-1566-4646}
\affiliation{
	\institution{Google and Johannes Kepler University Linz}
	\country{Austria}
}
\email{rmayrhofer@google.com}

\author{Jeffrey Vander Stoep}
\affiliation{
	\institution{Google}
	\country{Switzerland}
}
\email{jeffv@google.com}

\author{Chad Brubaker}
\affiliation{
	\institution{Independent}
	\country{USA}
}
\email{chad.m.brubaker@gmail.com}

\author{Dianne Hackborn}
\affiliation{
	\institution{Google}
	\country{USA}
}
\email{hackbod@google.com}
\author{Bram Bonné}
\affiliation{
	\institution{Google}
	\country{Switzerland}
}
\email{brambonne@google.com}

\author{G{\"u}liz Seray Tuncay}
\affiliation{
	\institution{Google}
	\country{USA}
}
\email{gulizseray@google.com}

\author{Roger Piqueras Jover}
\affiliation{
	\institution{Google}
	\country{USA}
}
\email{rogerpiqueras@google.com}

\author{Michael A. Specter}
\affiliation{
	\institution{Google}
	\country{USA}
}
\email{mikespecter@google.com}

\begin{abstract}
Android is the most widely deployed end-user focused operating system. 
With its growing set of use cases encompassing communication, navigation, media consumption, entertainment, finance, health, and access to sensors, actuators, cameras, or microphones, its underlying security model needs to address a host of practical threats in a wide variety of scenarios while being useful to non-security experts. 
To support this flexibility, Android's security model must strike a difficult balance between security, privacy, and usability for end users; provide assurances for app developers; and maintain system performance under tight hardware constraints. 
This paper aims to both document the assumed threat model and discuss its implications, with a focus on the ecosystem context in which Android exists. 
We analyze how different security measures in past and current Android implementations work together to mitigate these threats, and, where there are special cases in applying the security model in practice; we discuss these deliberate deviations and examine their impact.
\end{abstract}

\maketitle

\section{Introduction}
\label{sec:introduction}
Android is the most widely deployed end-user operating system. With more than three~billion monthly active devices~\cite{android-enterprise-whitepaper-2023} and a general trend towards mobile use of Internet services, Android is now the most common interface for global users to interact with digital services. Across different form factors, including phones, tablets, wearables, TVs, Internet-of-Things, automobiles, and more, Android supports a vast -- and still growing -- range of use cases including communication, media consumption, entertainment, finance, health, and physical sensors/actuators. 

Many of these applications are increasingly security and privacy critical, and Android's success continues to depend on the operating system's ability to provide sufficient assurances to all relevant stakeholders. 
Securing Android therefore requires balancing the different (and sometimes conflicting) needs of users, application developers, content producers, service providers, and employers. 

To manage the complexity of the diverse interests of all stakeholders, Android's security design has fundamentally been based on a \textbf{multi-party authorization model}: \emph{an action should only happen if all involved parties authorize it}.\footnote{Throughout the paper, the term ``authorization'' is used to refer to various technical methods of declaring or enforcing a party's intent, rather than the legal requirement of ``consent'' found in many privacy legal regimes around the world. In previous versions, the term ``consent'' was also used for the technical declaration of intent. This version clarifies the distinction by using the more technical notion of ``authorization''.} 
The rationale for this model is straightforward; If any party does not agree to an action, the safe-by-default choice is for that operation to be blocked. 
This is a significant departure from the security models implemented by traditional operating systems, which are focused on user access control and do not explicitly consider other stakeholders. 

 The goal of this paper is to document the Android security model, development, and implementation over time, with special consideration to the Android ecosystem's constraints and unique history.
While the multi-party authorization model has implicitly informed the architecture and design of the Android platform from the beginning, it has been refined and extended based on experience gathered from subsequent releases. 
Our hope is that this work will serve as a canonical resource on Android's security posture, providing insight for both researchers and the general public. 

\vspace{.5em}

\noindent Specifically, we make the following contributions:

\begin{enumerate}
	\item We provide security-relevant background on Android's history, design, and governance structure.
	
	\item We motivate and define the Android security model based on security principles and the wider context in which Android operates. Note that the core multi-party authorization model described in this paper has been implicitly informing Android security mechanisms since the earliest versions, and we therefore systematize knowledge that has, in part, been discussed only informally as folklore. 
 	
	\item We present Android's threat model, and discuss how the platform security model addresses commonly seen threats, including necessary special case handling. 
	
	\item We explain how the Android Open Source Project (AOSP), as the reference implementation of the Android platform, enforces the security model based on multiple interacting security measures at various layers of Android's stack.

	\item We identify open gaps and potential for future improvement of this implementation, as well as a number of open problems in Android security.
\end{enumerate}

\paragraph{Structure} We begin by introducing the ecosystem context and threat analysis that are the basis of the Android security model (Section~\ref{sec:background}). Next, we define the central security model (Section~\ref{sec:android-security-model}) and its implementation in the form of architecture and enforcement mechanisms on various layers of the operating system (Section~\ref{sec:implementation}). 
Finally, we discuss special cases (Section~\ref{sec:special-cases}) and basic related academic work 
 on Android security (Section~\ref{sec:related-work}).

All historical presentations are based on an analysis of security relevant changes to the whole AOSP code base between Android releases 4.x and 14 (inclusive), spanning about 13 years of code evolution.
Note that all implementation specific sections refer to Android 14 at the time of its initial release unless mentioned otherwise (cf.~\cite{androidblogpost-cellular-hardening,androidblogpost-cellular-security}, \cite{androidblogpost-security-privacy-13,androidblogpost-rust-2022}, \cite{androidblogpost-android-privacy-12,androidblogpost-android-enterprise-12}, \cite{androidblogpost-memory-autoinit-2020,androidblogpost-biometrics-2020,android-privacy-changes-11}, \cite{url-android-10-security-enhancements}, and~\cite{androidblogpost-pie-summary} for some of the relevant changes in Android 14 to 9, respectively). 
We will refer to earlier Android version numbers instead of their code names: 4.1--4.3 (Jelly Bean), 4.4 (KitKat), 5.x (Lollipop), 6.x (Marshmallow), 7.x (Nougat), 8.x (Oreo), and 9.x (Pie).
\section{Android background and scope}
\label{sec:background}

At its core, Android is an \emph{end-user focused operating system}, intended for everyday users of varying levels of expertise on consumer devices. As a result, it must be both useful to users and attractive to developers; user interfaces and workflows need to be safe by default and require explicit intent for any actions that could compromise security or privacy. This also means that the OS must not offload technically-challenging security or privacy decisions to non-expert users who are not sufficiently skilled or experienced to make them~\cite{Adams:1999:UE:322796.322806}. For developers, the operating system must provide adequate flexibility for a variety of use cases while providing sufficient guardrails to allow for interoperability between devices and protect against unintentional misuse of OS mechanisms. 

Here we provide an overview of security-relevant parts of the Android platform, particularly those that have direct implications to the security model. We focus on the actors involved, their various levels of control over the ecosystem, and how they interoperate in the development of Android devices in practice. We begin with a discussion of the Android ecosystem, how Android is defined by \emph{compatibility requirements}, and discuss the scope of this paper. Finally, we conclude the section with a discussion of the threats Android is designed to prevent.

\subsection{Android as an open ecosystem}
Android is not a vertically-integrated product, developed and maintained by one company. 
Instead, devices in the Android ecosystem are supported by hundreds of different Original Equipment Manufacturers (OEMs), who have collectively launched tens of thousands of Android devices in varying form factors~\cite{googleblogpost-android-choice-2018} (including smartphones, tablets, watches, glasses, headsets, Internet of things devices, handheld scanners/displays and other special-purpose devices, TVs, cars, and exercise equipment). Some OEMs do not have detailed technical expertise, and instead rely on Original Device Manufacturers (ODMs) for developing hardware and firmware and then re-package or re-label devices with their own brand. OEMs and ODMs also often depend on external silicon manufacturers to develop particular low-level hardware components for their devices, as well as firmware or kernel-level driver software. For simplicity, we often refer to these groups collectively as \emph{device manufacturers}.

While some parts of the platform may be customized or proprietary for different device manufacturers, AOSP provides \emph{reference implementations} for nearly all components. This includes the Linux kernel,\footnote{\url{https://android.googlesource.com/kernel/common/}} an ARM trusted execution environment called Trusty,\footnote{\url{https://android.googlesource.com/trusty/vendor/google/aosp/}} and a bootloader called \texttt{libavb} that provides a cryptographically verified boot process. 
In practice, device manufacturers often alter these reference implementations to form the distribution of Android that ultimately reaches the user's device. As a result, device manufacturers are responsible for customizing, building, distributing, and supporting Android for their devices. 

Another goal of Android is to provide application developers with maximum flexibility. As a result, Android:
\begin{enumerate}
    \item Explicitly supports installation of apps from arbitrary sources, including alternative app stores and apps independently distributed without a store at all.\footnote{Consequently, there is a long tail of apps with a very specific purpose, being installed on only a few devices, and/or targeting old Android API releases. The definition of and changes to APIs, therefore, need to consider the large number of such \emph{legacy} applications that are part of the Android ecosystem.}
    \item Allows applications to be written in any programming language, with or without runtime support, compiled or interpreted.\footnote{Android does not currently support non-Java language APIs for the basic process lifecycle control, because they would have to be supported in parallel, making the framework more complex and therefore more error-prone. Note that this restriction is not directly limiting, but apps need to have at least a small Java language wrapper to start their initial process and interface with fundamental OS services. }
\end{enumerate}
These design decisions have immediate implications for the security posture; Android cannot rely on compile-time checks or any other assumptions on the build environment, nor can it depend exclusively on a centralized store to enforce compatibility and security. 

\subsection{Android is defined by compatibility requirements}
Managing the complex ecosystem of device manufacturers, application developers, and users is a core challenge for Android. 
For example, without compatibility between devices, app developers would be forced to handle device-specific %
idiosyncracies, significantly increasing development time and effort. 
New device manufacturers would also have difficulty entering the market if applications were by-default incompatible with their systems, requiring app developers to spend resources to support their system, and users would have to contend with a complicated environment where certain applications may or may not function depending on the device.
Android succeeds in managing this complexity (and preventing these headaches) through various levels of compatibility requirements, standards, and compliance tests, allowing all participants to enjoy the network effects of the ecosystem.

In fact, the \emph{Android platform} is defined as the set of AOSP components that together form an operating system that adheres to standards called the Compatibility Definition Document (CDD)~\cite{url-cdd}. AOSP provides a series of tests, including the Compatibility Test Suite (CTS), which allow a device manufacturer to test if their system complies with the CDD. It may be the case that a system that passes CTS does not follow the CDD, for example the CDD is a ``whole device standard,’’ so a virtual machine or emulator running AOSP might pass CTS, but still not follow the CDD.

Devices that do not conform to the CDD (and/or do not pass CTS), by definition, are not Android. 
Further, devices that advertise themselves as Android as a trademarked name need to, at minimum, pass CTS and an additional set of tests called the Vendor Test Suite (VTS). 
Finally, Android devices are subjected to additional requirements to be certified to run Google Mobile Services, which we call GMS Android. 
Though devices based purely on AOSP are only required to ensure that they meet publicly documented compatibility tests, those shipping with Google services are held to a stricter standard, including additional compliance tests focusing on security and privacy. 

\begin{table*}[h]
    \centering
    \begin{tabular}{l|cccc|c}
        \toprule
                                   &  CDD & CTS & VTS & GMS \& Security Tests & Is Android?\\
        \midrule
        Untested fork of AOSP     &    \xmark    & \xmark  &      \xmark       &    \xmark       & \xmark \\
        AOSP                       &  \cmark & \cmark &  \xmark  &   \xmark    & \cmark \\
        Trademarked Android        &  \cmark &  \cmark &  \cmark &    \xmark     & \cmark \\
        GMS Android                &  \cmark &  \cmark & \cmark  &\cmark & \cmark \\
        \bottomrule
    \end{tabular}
    
    \caption{The various types of ``Android'' systems. A \cmark \ here indicates that the OS is guaranteed, in some way, to be compliant with the requirement, whereas \xmark \ indicates that the OS is either not compliant or that the compliance is not guaranteed. For example, something forked from AOSP and not tested may happen to comply with the Android specification, but this cannot be guaranteed  without passing the CDD and the CTS.}
    \label{table:attacks}
\end{table*}

The term Android is commonly used to refer to a number of different items in the Android ecosystem, including:
\begin{itemize}
    \item The Android Open Source Project (AOSP), which is a reference implementation of the Android operating system.
    \item Operating systems forked from AOSP, that happen to be capable of running Android applications.
    \item Devices running trademarked Android operating system.
    \item Devices running trademarked Android, as well as Google services.
\end{itemize}

This paper focuses on security and privacy measures in the Android platform itself, i.e.\ code running on user devices that is part of AOSP. 
We define the \emph{platform} as the set of AOSP components that together form a complete system that is compliant with CDD, much of which is enforced by a suite of tests defined by Google called the Compatibility Test Suite or CTS (more fully discussed below in Section~\ref{sec:android-security-model}).

In contrast, there are a number of proprietary services that, while they do affect the security posture of Android in practice, are not relevant for understanding the security model we outline in this paper. These include specific code running under Private Compute Core (see Section~\ref{subsec:app-pcc}), Google Mobile Services (GMS), Google Play Services, the Google Play Store, Google Search, Chrome, and other standard apps that are colloquially considered part of Android, as they provide dependencies for common services such as location estimation or cloud push messaging. Android devices certified to support GMS are publicly listed.\footnote{\url{https://storage.googleapis.com/play_public/supported_devices.html}} 
While replacements for these components exist (including an independent, minimal open source version called \texttt{microG}\footnote{\url{https://github.com/microg/android_packages_apps_GmsCore/wiki}}), they may be incomplete or behave differently than what is described in this work. 
Again, we do not consider these services as part of the platform, as they are also subject to the security policy defined and enforced by AOSP components.\footnote{In terms of higher-level security measures, there are services complementary to those implemented in AOSP in the form of Google Play Protect scanning applications submitted to Google Play and on-device (Verify Apps or Safe Browsing as opt-in services) as well as Google Play policy and other legal frameworks.
These are also out of scope of the current paper, but are covered by related work~\cite{androidblogpost-safe-browsing-2018,androidblogpost-malicious-apps-2018,androidblogpost-play-p2p-distribution-2018,androidblogpost-play-machine-learning-2018}. However, it is worth noting that restrictions in Google Play can work in tandem with the platform to yield significant positive effects for security. For example, Play now requires that apps target a recent Android API level, which will allow the Android platform to deprecate and remove APIs known to be abused or that have had security issues in the past~\cite{androidblogpost-play-api-updates-2017}. Similarly, Google Play Services have been used to ``backport'' security improvements (such as permission revocation~\cite{androidblogpost-permission-revocation}) to devices running older versions of AOSP.} 

There are also a class of devices that, though they may run a fork of AOSP, do not adhere to CTS/CDD or follow the threat model described here, and so we cannot consider or adequately describe the security of these devices – these ``forked'' platforms are outside the scope of this work. In a practical sense, it is difficult to reason about their threat model, and any customizations done to these systems may not be compatible with the security and privacy goals of Android.

\subsection{Threat model}
\label{subsec:threat-model}
Threat models for mobile devices are extended from those commonly used for desktop or server operating systems for two major reasons: by definition, mobile devices are easily lost or stolen, and they connect to untrusted networks as part of their expected usage. At the same time, by being close to users at most times, they are also exposed to more privacy sensitive data than many other categories of devices. Therefore, modern mobile device platforms, among other extensions, primarily tend to assign less trust to locally installed applications than traditional desktop operating systems, which may be argued to not have sufficiently adapted to the changed threat landscape.\footnote{Newer (Linux) desktop operating systems have started adopting some of the concepts like read-only system images and sandboxed applications as well, although the basic association of permissions to the logged-in user account is still problematic under the threat of actively malicious local applications.} Recent work~\cite{art-wiley-scn-2014} previously introduced a layered threat model for mobile devices which we adopt for discussing the Android security model within the scope of this paper; however, where meaningful, we order threats in each category with lower numbers representing more constrained and higher numbers more capable adversarial settings:

\paragraph{Adversaries can get physical access to Android devices.}
For all mobile and wearable devices, we have to assume that they will potentially fall under physical control of adversaries at some point. The same is true for other Android form factors such as wearables, cars, TVs, etc. Therefore, we assume Android devices to be either directly accessible to adversaries or to be in physical proximity to adversaries as an explicit part of the threat model. This includes loss or theft, but also multiple (benign but potentially curious) users sharing a device (such as a TV or tablet). We derive specific threats due to \emph{physical} or \emph{proximal (P)} access:
\begin{description}
	\item[T.P1] (Screen locked or unlocked) devices in physical proximity to, but not under direct control of, an adversary with the assumed capability to control all available radio communication channels, including cellular, WiFi, Bluetooth, UWB, GPS, NFC, and FM, e.g.\ direct attacks through Bluetooth~\cite{blueborne,DBLP:conf/wisec/ClassenH19}. Although NFC could be considered to be a separate category to other proximal radio attacks because of the scale of distance, we still include it in the threat class of proximity instead of physical control.
	\item[T.P2] Powered-off devices under complete physical control of an adversary (with potentially high sophistication up to nation state level attackers), e.g.\ border control or customs checks.
	\item[T.P3] Screen locked devices under complete physical control of an adversary, e.g.\ thieves trying to exfiltrate data for additional identity theft.
	\item[T.P4] Screen unlocked (shared) devices under control of an authorized but different user, e.g.\ intimate partner abuse, voluntary submission to a border control or customs check.
\end{description}

\paragraph{Network communication is untrusted.}
The standard assumption of network communication under complete control of an adversary certainly also holds for Android devices. This includes the first hop of network communication (e.g.\ captive WiFi portals breaking TLS connections and malicious fake access points) as well as other points of control (e.g.\ mobile network operators or national firewalls), summarized in the usual Dolev-Yao model~\cite{art-dolev-yao} 
with additional relay threats for short-range radios (e.g.\ NFC or BLE wormhole attacks~\cite{Roland2013NFC}). For practical purposes, we mainly consider three \emph{network-level (N)} threats:
\begin{description}
	\item[T.N1] Passive eavesdropping and traffic analysis, including tracking devices within or across networks, e.g.\ based on MAC address or other device network identifiers.
	\item[T.N2] Active manipulation of network traffic, e.g.\ machine-in-the-middle (MITM) or on-path attacks (OPA) on TLS connections or relaying.
	\item[T.N3] Adversarial \emph{cellular} network provider, e.g.\ rogue cellular operator~\cite{intercept_iran_carriers} or rogue/false cellular base station~\cite{practicalattacksLTE,EFF_gotta_catch_em_all}. The peculiarities of cellular protocols expose devices to threats that are unique to these types of networks and not addressed by mitigations to [T.N1] and [T.N2]~\cite{androidblogpost-cellular-security}.
\end{description}
These threats are different from [T.P1] (proximal radio attacks) in terms of scalability of attacks. Controlling a single choke point in a major network can be used to attack a large number of devices, while proximal (i.e., last hop) radio attacks require physical proximity to target devices.

\paragraph{Untrusted code is executed on the device.}
One fundamental difference to other mobile operating systems is that Android intentionally allows, with explicit authorization by end users, installation of \emph{application (A) code} from arbitrary sources, and does not enforce vetting of apps by a central instance. This implies attack vectors on multiple levels (cf.~\cite{art-wiley-scn-2014}):
\begin{description}
	\item[T.A1] Abusing APIs supported by the OS with malicious intent, e.g.\ spyware.
	\item[T.A2] Abusing APIs provided by other apps installed on the device~\cite{SVE-2018-11633}.
	\item[T.A3] Untrusted code from the web (i.e.\ JavaScript) is executed without explicit authorization~\cite{luo2011attacks, chin2014bifocals, tuncay2016draco}. 
	\item[T.A4] Mimicking system or other app user interfaces to confuse users (based on the knowledge that standard in-band security indicators are not effective~\cite{Dhamija:2006:WPW:1124772.1124861,art-puc2015-security-zones}), e.g.\ to input PIN/password into a malicious app~\cite{fernandes_android_2016}, permission phishing~\cite{tuncay2020see}.
	\item[T.A5] Reading content from system or other app user interfaces, e.g.\ to screen-scrape confidential data from another app~\cite{Jang:2014:AAE:2660267.2660295,kraunelis14-accessibility}.
	\item[T.A6] Injecting input events into system or other app user interfaces~\cite{fratantonio17:cloakdagger}.
	\item[T.A7] Exploiting bugs in the OS to escalate privileges or gain code execution, e.g.\ in the kernel, drivers, or system services~\cite{CVE-2017-13177,SVE-2018-11599,CVE-2018-9341,CVE-2017-17558}, or in the firmware running on other processors within the SoC that perform various specialized tasks (e.g.\ cellular communications)~\cite{grassi_BH2021_RCEbaseband}, or in the protected components of other apps~\cite{tuncay2018resolving, li2021android}.
	\item[T.A8] Surreptitiously adding potentially harmful code to system images or other code or data executed or interpreted on-device through insider capabilities such as access to private code signing keys.
\end{description}

\paragraph{Untrusted content is processed by the device.}
In addition to directly executing untrusted code, devices process a wide variety of untrusted data, including rich (in the sense of complex structure) media. This directly leads to threats concerning the processing of \emph{data (D)} and metadata:
\begin{description}
	\item[T.D1] Abusing unique identifiers for targeted attacks (which can happen even on trusted networks), e.g.\ using a phone number or email address for spamming or correlation with other datasets, including locations. This includes using stable device identifiers to cross profile boundaries or factory resets, without explicit involvement from the platform.
	\item[T.D2] Exploiting code that processes untrusted content in the OS, firmware, or apps, e.g.\ in media libraries~\cite{stagefright-vulnreport}, or libraries parsing ASN.1-encoded cellular messages~\cite{baseband_unisoc}. This can be both a local as well as a remote attack surface, depending on where input data is taken from.
\end{description}

\paragraph{Many stakeholders in the ecosystem can act as supply chain attack vectors.}
In addition to the explicit [T.A8] modeling injection of harmful code shipped with the device or apps, (technical) \emph{insider attacks} can occur at many more levels in the complex supply chain of hard- and software vendors, including chipset manufacturers, ODMs, OEMs, contributors to the AOSP code base, third party libraries, or even malicious insiders at the platform vendor (i.e., Google). Insiders with privileged access may be able to leak or abuse access to code signing keys, directly modify code shipped to all users of a particular subsystem or device model, create targeted modifications for only a subset of users, or tamper with keys created for individual user devices (e.g., to inject weak cryptographic keys or properties useful for fingerprinting in the field). Android is far more concerned about attacks on integrity than on confidentiality of code running on user devices, particularly considering that most of the code is open source in the first place. 

Another class of supply chain attacks are \emph{organizational attacks} on a legal or political level. These may for example take the form of compelled technical insider attacks to access available data or make changes to code controlled by an organization, or targeted or national/global censorship of code or data available or accessed through a platform. We note that, due to the open nature of the Android ecosystem without a single point of vetting of code or data, it is inherently more robust to censorship or organizational attacks against single targets.

The possibility of insider and/or organizational attacks at many levels is an effect of the ecosystem size, and such attacks need to be part of a realistic threat model.

\section{The Android Platform Security Model}
\label{sec:android-security-model}
The basic security model described in this section has informed the design of Android, and has been refined but not fundamentally changed. Given the ecosystem context and threat model explained above, the Android security model balances security and privacy requirements of users with security requirements of applications and the platform itself. The threat model described above includes threats to all stakeholders, and the security model and its enforcement by the Android platform aims to address all of them.
The Android platform security model is informally defined by five rules:

\paragraph{\circled{1} Multi-party authorization.}\label{rule:1}
No action should be executed unless all main parties agree --- in the standard case, these are \emph{user}, \emph{platform}, and \emph{developer} (implicitly representing stakeholders such as content producers and service providers). Any one party can veto the action. This multi-party authorization spans the traditional two dimensions of subjects (i.e., users and application processes) vs.\ objects (i.e., files, network sockets and IPC interfaces, memory regions, virtual data providers, etc.) that underlie most security models (e.g.~\cite{Tanenbaum:2014:MOS:2655363}). Any party (or more generally actor) that creates a data item is implicitly granted control over this particular instance of data representation. Focusing on (regular and pseudo) files as the main category of objects to protect, the default control over these files depends on their location and which party created them:
\begin{itemize}
	\item Data in shared storage is controlled by users.
	\item Data in private app directories and app virtual address space is controlled by apps.
	\item Data in special system locations is controlled by the platform (e.g.\ list of granted permissions).
\end{itemize}
Data in run-time memory (RAM) is by default controlled by the respective platform or app process. However, it is important to point out that, under multi-party authorization, even if one party primarily controls a data item, it may only act on it if the other involved parties authorize the respective access. \emph{Control over data also does not imply ownership} --- which is a legal concept rather than a technical one and therefore outside the scope of an OS security model. 

The same notion of authorization applies to data or resources that are technically controlled by external parties outside the Android platform, such as cloud services. Any external dependencies of apps or the platform itself are treated as parts of the respective party from an authorization point of view: services provided by the OEM are considered to be part of the platform as long as they obey the security model, and services used by apps are considered to be part of the app domain (e.g., external web resources loaded by an embedded WebView). Data on such external services may only be used in ways covered by mutual authorization --- both for data retrieved from external services and used on-device and data created on-device and then used externally.

While this principle has long been the default for filesystem access control (DAC, cf.\ Section~\ref{subsec:permissions} below), we consider it a global model rule and exceptions such as device backup (cf.\ Section~\ref{sec:special-cases}) can be argued about within the scope of the security model. There are other corner cases in which only a subset of all parties may need to authorize --- for actions in which the user only uses platform/OS services without involvement of additional apps --- or an additional party may be introduced (e.g.\ on devices or profiles controlled by a mobile device management, this policy is also considered as a party for authorizing an action).

\begin{leftbar}
	\emph{Public} information and resources are out of scope of this access control and available to all parties; particularly all static code and data contained in the AOSP system image and apps (mostly in the Android Package (APK) format) is considered to be public (cf.\ Kerckhoff's principle) --- if an actor publishes the code, this is interpreted as implicit consent to access. However, it is generally accepted that such public code and data is read-only to all parties and its integrity needs to be protected, which is explicitly in scope of the security measures.
\end{leftbar}

\paragraph{\circled{2} Open ecosystem access.}\label{rule:2}
Both users and developers are part of an open ecosystem that is not limited to a single application store. Central vetting of developers or registration of users is not required. This aspect has an important implication for the security model: generic app-to-app interaction is explicitly supported. Instead of creating specific platform APIs for every conceivable workflow, app developers are free to define their own APIs they offer to other apps.

\paragraph{\circled{3} Security is a compatibility requirement.}\label{rule:3}
The security model is part of the Android specification, which is defined in the Compatibility Definition Document (CDD)~\cite{url-cdd} and enforced by the Compatibility (CTS), Vendor (VTS), and other test suites. Devices that do not conform to CDD and do not pass CTS are not Android. Within the scope of this paper, we define \emph{rooting} as modifying the system to allow starting processes that are not subject to sandboxing and isolation. Such rooting, both intentional and malicious, is a specific example of a non-compliant change which violates CDD. As such, only CDD-compliant devices are considered. While many devices support unlocking their bootloader and flashing modified firmware\footnote{Google Nexus and Pixel devices as well as many others support the standard \texttt{fastboot oem unlock} command to allow flashing any firmware images to actively support developers and power users. However, executing this unlocking workflow will forcibly factory reset the device, wiping all data, to make sure that security guarantees are not retroactively violated for data on the device.}, such modifications may be considered incompatible under CDD if security assurances do not hold. Verified boot and hardware key attestation can be used to validate if currently running firmware is in a known-good state, and in turn may influence consent decisions by users and developers.

\paragraph{\circled{4} Factory reset restores the device to a safe state.}\label{rule:4}
In the event of security model bypass leading to a persistent compromise, a factory reset, which wipes/reformats the writable data partitions, returns a device to a state that depends only on integrity protected partitions. In other words, system software does not need to be re-installed, but wiping the data partition(s) will return a device to its default state. Note that the general expectation is that the read-only device software may have been updated since originally taking it out of the box, which is intentionally not downgraded by factory reset. Therefore, more specifically, factory reset returns an Android device to a state that only depends on system code that is covered by \emph{Verified Boot}, but does not depend on writable data partitions. An important aspect is supply chain security of these factory images in the sense of integrity and authenticity of system code (cf.\ Section~\ref{subsec:integrity-gbt}).

\paragraph{\circled{5} Applications are security principals.}\label{rule:5}
The main difference to traditional operating systems that run apps in the context of the logged-in user account is that Android apps are not considered to be fully authorized agents for user actions. In the traditional model typically implemented by server and desktop OSes, there is often no need to even exploit the security boundary because running malicious code with the full permissions of the main user is sufficient for abuse. Examples are many, including file encrypting ransomware~\cite{7536529,10.1007/978-3-319-20550-2_1}, which does not violate the OS security model if it simply re-writes all the files the current user account has access to, and private data leakage (e.g.\ browser login tokens~\cite%
{rfc6819-oauth-security}, history or other tracking data, cryptocurrency wallet keys, etc.).

\bigskip

\paragraph{Summary}
Even though, at first glance, the Android security model grants less power to users compared to traditional operating systems that do not impose a multi-party consent model, there is an immediate benefit to end users: if one app cannot act with full user privileges, the user cannot be tricked into letting it access data controlled by other apps. In other words, requiring application developer consent -- enforced by the platform -- helps avoid user confusion attacks and therefore better protects private data.

The Android platform security model does not currently have a simple, consistent representation in formal notation because these rules evolved from practical experience instead of a top-down theoretical design; the meaning of the term ``model'' is consequently slightly different from how conventional security models use it. Balancing the different requirements of a complex ecosystem is a large scale engineering problem that requires layers of abstraction. Therefore, we have to combine multiple different security controls, such as memory isolation, filesystem DAC/MAC, biometric user authentication, or network traffic encryption, that operate under their own respective models and are not necessarily consistent with each other (see e.g.\ \cite{247662} for interactions between only the DAC and MAC policies). The five rules are, at the time of this writing, the simplest expression of how these different security controls combine at the meta level.

Appendix~\ref{appendix:formal-rules} gives a first, albeit incomplete, formalization of the access control properties of these rules. It is subject to future work to model all important aspects more formally and to reason about the cross-abstraction interactions of these rules with lower level models of underlying security controls.

\section{Implementation}
\label{sec:implementation}
Android's security measures implement the security model and are designed to address the threats outlined above. A high-level summary of the historical design and development can be found in the 2023 Android Enterprise Security White Paper~\cite{android-enterprise-whitepaper-2023}, including the fundamental use of Linux kernel user separation mechanisms as one of the underlying security building blocks in Android. In this section we describe the combination of multiple interlocking security measures and indicate which threats they mitigate, taking into account the architectural security principles of ``defense in depth'' and ``safe by design'':

\paragraph{Defense in depth.}
A robust security system is not sufficient if the acceptable behavior of the operating system allows an attacker to accomplish their goals without bypassing the security model (e.g.\ ransomware encrypting all files it has access to under the access control model). Specifically, violating any of the above principles should require such bypassing of controls on-device, in contrast to relying on off-device verification (e.g.\ at build time).

Therefore, the primary goal of any security system is to enforce its model. For Android operating in a multitude of environments (see above for the threat model), this implies an approach that does not immediately fail when a single assumption is violated or a single implementation bug is found, even if the device is not up to date.  Defense in depth is characterized by rendering individual vulnerabilities more difficult or impossible to exploit, and increasing the number of vulnerabilities required for an attacker to achieve their goals. We primarily adopt four common security strategies to prevent adversaries from bypassing the security model: \emph{isolation and containment} (Section~\ref{subec:isolation-and-containment}), \emph{exploit mitigation} (Section~\ref{subsec:exploit-mitigation}), \emph{integrity} (Section~\ref{subsec:integrity}), and \emph{patching/updates} (Section~\ref{subsec:patching}), as well as the special case of \emph{cellular security} defense mechanisms for mobile device networks (Section~\ref{subec:cellular-network-security}). 

\paragraph{Safe by design/default.}
Components should be safe by design. That is, the default use of an operating system component or service should always protect security and privacy assumptions, potentially at the cost of blocking some use cases. This principle applies to modules, APIs, communication channels, and generally to interfaces of all kinds. When variants of such interfaces are offered for more flexibility (e.g.\ a second interface method with more parameters to override default behavior), these should be hard to abuse, either unintentionally or intentionally. Note that this architectural principle targets developers, which includes device manufacturers, but implicitly includes users in how security is designed and presented in user interfaces.
Android targets a wide range of developers and intentionally keeps barriers to entry low for app development. Making it hard to abuse APIs not only guards against malicious adversaries, but also mitigates genuine errors resulting e.g.\ from incomplete knowledge of an interface definition or caused by developers lacking experience in secure system design. As in the defense in depth approach, there is no single solution to making a system safe by design. Instead, this is considered a guiding principle for defining new interfaces and refining -- or, when necessary, deprecating and removing -- existing ones. For guarding user data, the basic strategies for supporting safety by default are: \emph{enforced authorization} (Section~\ref{subsec:authorization}), \emph{user authentication} (Section~\ref{subsec:lock-screen}), and \emph{by-default encryption at rest} (Section~\ref{subec:encryption-at-rest}) and \emph{in transit} (Section~\ref{subec:encryption-in-transit}).

\subsection{Enforcing meaningful consent in authorization decisions}
\label{subsec:authorization}
Methods allowing for meaningful consent vary greatly depending on the actor and situational constraints. 

\begin{leftbar}
	We use three examples to better describe the authorization parties:
	\begin{itemize}
            \item An app getting access to the user's location requires:
            \begin{itemize}
                \item app authorization by specifying in its manifest that it may request this permission (for auditability), and at runtime in an appropriate context telling the platform it would like to request this permission;
                \item platform authorization by verifying the app has correctly specified the permission in its manifest, and prompting the user to allow this access; and
                \item user authorization by affirming in the runtime permission prompt that access is allowed.
            \end{itemize}
            
            \item Opening an external file for an app to access requires:
            \begin{itemize}
                \item app authorization by using \texttt{ACTION\_GET\_CONTENT} to request that it get access to a file;
                \item platform authorization by presenting the available files and creating and enforcing the accessibility of only the appropriate file; and
                \item user authorization by using the platform's \texttt{ACTION\_GET\_CONTENT} UI to select the file(s) they would like the app to access.
            \end{itemize}
            
            \item Sharing data from app $\mathbf{A}$ to app $\mathbf{B}$ requires:
		\begin{itemize}
			\item user authorization by selecting data in app $\mathbf{A}$ that they would like to share;
			\item app $\mathbf{A}$ authorization by using \texttt{ACTION\_SEND} to request that this data be shared elsewhere; 
                \item platform facilitates and enforces the share by presenting a list of possible share targets (if there is ambiguity) and tracking the access to this data, which forms a temporary trust relationship between the two apps;
			\item user authorization by reviewing and approving app $\mathbf{B}$'s UI of the data being shared to it (such as a new message containing the data that will be sent); and
                \item app $\mathbf{B}$ authorization by accepting and processing the shared data.
		\end{itemize}
	\end{itemize}
\end{leftbar}

Actors authorizing any action must be empowered to base their decision on information about the action and its implications and must have meaningful ways to grant or deny this authorization. This applies to both users and developers, although very different technical means of enforcing (lack of) authorization apply. For users, the technical mechanism of authorization is a means of providing consent for an action. Authorization is not only required from the actor that created a data item, but from all involved actors. Authorization decisions should be enforced and not self-policed, which can happen at run-time (often, but not always, through platform mediation) or build respectively distribution time (e.g.\ developers including or not including code in particular app versions).

\subsubsection{Developer(s)}
Unlike traditional desktop operating systems, Android ensures that the developer authorizes actions on their app or their app's data. This prevents large classes of abusive behavior where unrelated apps inject code into or access/leak data from other applications on a user's device.

Authorization for developers is given via the code they sign and the system executes, uploading the app to an app store and agreeing to the associated terms of service, and obeying other relevant policies (such as CDD for code by an OEM in the system image). For example, an app can authorize the user sharing its data by providing a respective mechanism, e.g.\ based on OS sharing methods such as built-in implicit \texttt{Intent} resolution chooser dialogs~\cite{url-intent-filters} and by utilizing custom (e.g., signature) permissions to regulate access to the exported app components~\cite{android-custom-permissions}. Another example is debugging; as assigned virtual memory content is controlled by the app, debugging from an external process is only allowed if an app authorizes it (specifically through the \texttt{debuggable} flag in the app manifest). By uploading an app to the relevant app store, developers also provide the authorization for this app to be installed on devices that fetch from that store under appropriate preconditions (e.g.\ after successful payment).

For authorization to be clear from a developer point of view, the platform needs to ensure that APIs and their behaviors are clear and the developer understands how their application is interacting with or providing data to other components. We assume that developers of varying skill levels may not have a complete understanding of security nuances, and as a result APIs must also be safe by default and difficult to incorrectly use in order to avoid accidental security regressions. One example of a lesson learned in these regards is the changed default for exporting app components to \texttt{false}, away from the ``smart'' behavior of default \texttt{true} if the component has an intent-filter, which caused unintentional over-exporting in the past.

Android~9 introduced a major change by only supporting access to APIs explicitly listed as external\footnote{See \url{https://developer.android.com/reference/packages}} and putting restrictions on others~\cite{url-private-sdk-apis}. Developer support was added e.g.\ in the form of specific log messages to point out internal API usage for debuggable versions of apps. This has two main benefits: a) the attack surface is reduced, both towards the platform and apps that may rely on undefined and therefore changing internal behavior; and b) refactoring of internal platform interfaces and components from one version to another is enabled with fewer app compatibility constraints. 

In order to ensure that it is the app developer and not another party that is authorizing, applications are signed by the developer. This prevents third parties from replacing or removing code or resources in order to change the app's intended behavior\footnote{One exception is app stores holding those signing keys and signing a final version of the APK as shipped to end-user devices in the name of the developer.}. However, the app signing key is trusted implicitly upon first installation, so replacing or modifying apps in transit when a user first installs them (e.g.\ when initially side-loading apps) is currently out of scope of the platform security model. Previous Android versions relied on a single developer certificate that was trusted on initial install of an app and therefore made it impossible to change the underlying private key e.g.\ in the case of the key having become insecure~\cite{10.1145/2627393.2627397}. Starting with Android~9, independently developed key rotation functionality was added with APK Signature Scheme v3 \cite{url-apk-signing} to support delegating the ability to sign to a new key by using a key that was previously granted this ability by the app using so-called \emph{proof-of-rotation} structs\footnote{The Google Play app store now explicitly supports key rotation through Play Signing, but does not yet support key rotation with multiple developer-held keys. The Android platform itself supports arbitrarily complex key rotation strategies.}.

These three examples (i.e., default component exporting, controlled access to internal Android platform components and developer signing key rotation) highlight that handling multi-party authorization in a complex ecosystem is challenging even from the point of a single party: some developers may wish for maximum flexibility (i.e., access to all internal components and arbitrarily complex key handling), but the majority tends to be overwhelmed by the complexity. As the ecosystem develops, changes are therefore necessary to react to lessons learned. In these examples, platform changes largely enabled backwards compatibility without changing (i.e., no impact when key rotation is not used by a developer) or breaking (i.e., most apps do not rely on internal APIs) existing apps. When changes for developers are necessary, these need to be deployed over a longer period to allow adaptation, typically with warnings in one Android release and enforced restrictions only in the next one.

\subsubsection{The Platform}
While the platform, like the developer, authorizes via code signing, the goals are quite different: the platform acts to ensure that the system functions as intended. This includes enforcing regulatory or contractual requirements (e.g.\ communication in cellular networks) as well as taking an opinionated stance on what kinds of behaviors are acceptable (e.g.\ mitigating apps from applying deceptive behavior towards users). Platform authorization is enforced via Verified Boot (see below for details) protecting the system images from modification, internal compartmentalization and isolation between components, as well as platform applications using the platform signing key and associated permissions, much like applications.

\begin{leftbar}
	\paragraph{Note on the platform as a party:} Depending on how the involved stakeholders (i.e., parties for authorization) and enforcing mechanisms are designated, either an inherent or an apparent asymmetry of power to authorize may arise: 
	
	(a) If the Android ``platform'' is seen as a single entity, composed of hardware, firmware, OS kernel, system services, libraries, and app runtime, then it may be considered omniscient in the sense of having access to and effectively controlling all data and processes on the system. Under this point of view, the conflict of interest between being one party of authorization and simultaneously being the enforcing agent gives the platform overreaching power over all other parties.
	
	(b) If Android as a platform is considered in depth, it consists of many different components. These can be seen as individual representatives of the platform for a particular interaction involving multi-party authorization, while other components act as the enforcing mechanism for that authorization. In other words, the Android platform is structured in such a way as to minimize trust in itself and contain multiple mechanisms of isolating components from each other to enforce each other's limitations (cf.\ Section~\ref{subec:isolation-and-containment}). One example is playing media files: even when called by an app, a media codec cannot directly access the underlying resources if the user has not granted this through the media server, because MAC policies in the Linux kernel do no allow such bypass (cf.\ Section~\ref{subsec:system-process-sandbox}). Another example is storage of cryptographic keys, which is isolated even from the Linux kernel itself and enforced through hardware separation (cf.\ Section~\ref{subsec:hardware-sandbox}). While this idealized model of platform parties requiring authorization for their actions is the abstract goal of the security model we describe, in practice there still are individual components that sustain the asymmetry between the parties. Each new version of Android continues to further strengthen the boundaries of platform components among each other, as described in more detail below.
	
	Within the scope of this paper, we take the second perspective when it comes to notions of authorization involving the platform itself, i.e.\ considering the platform to be multiple parties whose authorization is being enforced by independent mechanisms (i.e., mostly the Linux kernel isolating platform components from each other, but also including out-of-kernel components in a trusted execution environment). However, when talking about the whole system implementing our Android security model, in favor of simpler expression we will generally refer to the platform as the combination of all (AOSP) components that together act as an enforcing mechanism for other parties, as defined in the introduction.
\end{leftbar}

Lessons learned over the evolution of the Android platform are clearly visible through the introduction of new security mitigations and tightening of existing controls, as summarized in Tables~\ref{tab:application-sandboxing-improvements} to~\ref{tab:network-security-improvements} and too extensive to describe here. The platform not only manages authorization for its own components, but mediates user and developer authorization responses, and therefore has to adapt to changes in the ecosystem (e.g.\ to prevent apps deceiving users to provide authorization against their wishes through use of deceptive strings, namespaces, and links).

\subsubsection{User(s)}
Achieving meaningful user authorization is by far the most difficult and nuanced challenge in determining meaningful consent. For users that use their Android devices with their own personal data and for their own, potentially highly sensitive purposes, technical \emph{authorization} of software actions directly means \emph{consent} to having their data processed or actions taken in their name. That is, especially for users --- in contrast to developers and the platform or other parties ---, this consent implies concerns beyond the pure technical mechanism of providing authorization decisions and reaches into the multi-faceted aspects of \emph{meaningful and informed consent} that includes legal, usability, social, and ethical concerns.

Some of the guiding principles concerning this user consent have always been core to Android, while others were refined based on experiences over 10\,years of development so far:
\begin{itemize}
	\item \textbf{Avoid over-prompting.} Over-prompting the user leads to prompt fatigue and blindness (cf.~\cite{brinton16-warning-to-wallpaper}). Prompting the user with a yes/no prompt for every action does not lead to meaningful consent as users become blind to the prompts due to their regularity.
	
	\item \textbf{Prompt in a way that is understandable.} Users are assumed not to be experts or understand nuanced security questions (cf.~\cite{Felt:2012:APU:2335356.2335360}). %
	Prompts and disclosures must be phrased in a way that a non-technical user can understand the effects of their decision.
	
	\item \textbf{Prefer pickers and transactional consent over wide granularity.} When possible, we limit access to specific items instead of the entire set through \texttt{ACTION\_GET\_CONTENT}, typically presenting the available files and creating and enforcing the accessibility of only the appropriate file. This also allows the user to select from a list of applications that can provide content of the requested type, such as apps providing access to cloud file storage. As a more specific example, the Photo Picker allows the user to select a specific picture to share with the application instead of using the Storage permission. These both limit the data exposed as well as present the choice to the user in a clear and intuitive way.
	
	\item \textbf{The OS must not offload a difficult problem onto the user.} Android regularly takes an opinionated stance on which behaviors are too risky to be allowed and may avoid adding functionality that may be useful to a power user but dangerous to an average user.
	
	\item \textbf{Provide users a way to undo previously made decisions.} Users can make mistakes. Even the most security and privacy-savvy users may simply press the wrong button from time to time, which is even more likely when they are being tired or distracted. To mitigate against such mistakes or the user simply changing their mind, it should be easy for the user to undo a previous decision whenever possible. This may vary from denying previously granted permissions to removing an app from the device entirely. It has always been a core design goal of the Android platform to allow an app to be cleanly removed without depending on it to do so, so that apps cannot --- through malicious behavior or simple coding errors --- prevent their own removal. This makes malware removal trivial, compared to often requiring a complete platform re-install on traditional desktop/server operating systems.
    
    Over time, users might forget about consent decisions made in the past, especially if an app hasn't been used for a long time. For this reason, since Android~11, the platform automatically revokes permissions when an app hasn't been used for some time~\cite{app-hibernation}.
\end{itemize} 

Additionally, it is critical to ensure that the user who is authorizing is the legitimate user of the device and not another person with physical access to the device ([T.P2]-[T.P4]), which directly relies on the next component in the form of the Android lockscreen. Implementing model \hyperref[rule:1]{rule \circled{1} (multi-party authorization)} is cross-cutting on all system layers.

For devices that do not have direct, regular user interaction (embedded IoT devices, shared devices in the infrastructure such as TVs, etc.), user authorization may be given slightly differently depending on the specific form factor. A smart phone may often act as a UI proxy to configure consent/policy for other embedded devices. For the remainder of this paper but without loss of generality, we primarily assume smart phone/tablet type form factors with direct user interaction.

As with developer authorization, lessons learned for user consent over the development of the ecosystem will require changes over time. The biggest changes for user autorization were the introduction of runtime permissions with Android~6.0, moving the point of authorization from install time to access time and therefore granting users better implicit context and granularity, and non-binary, explicitly context dependent permissions with Android~10 (cf.\ Section~\ref{subsec:permissions}). Other examples are restrictions to accessibility service APIs (which require user authorization but were abused), clipboard access and background activity launches starting in Android~10, app hibernation in Android~11~\cite{app-hibernation}, and mitigations against tapjacking attacks in Android~12 (cf.\ Table~\ref{tab:application-sandboxing-improvements}).

\subsection{Authentication}
\label{subsec:lock-screen}
Authentication is a gatekeeper function for ensuring that a system interacts with its owner or legitimate user. On mobile devices the primary means of authentication is via the lockscreen.
Note that a lockscreen is an obvious trade-off between security and usability: On the one hand, users unlock phones for short (10-250 seconds) interactions about 50 times per day on average and even up to 200 times in exceptional
cases~\cite{hintze2017deviceusage,Falaki_2010}, and the lockscreen is obviously an immediate hindrance to frictionless interaction with a device~\cite{paper-ubicomp2014adjunct-mobile-device-locking-usage,paper-momm2014-mobile-device-usage}. On the other hand, devices without a lockscreen are immediately open to being abused by unauthorized users ([T.P2]--[T.P4]), and the OS cannot reliably enforce user consent without authentication. 

In their current form, lockscreens on mobile devices largely enforce a binary model --- either the whole phone is accessible, or the majority of functions (especially all security or privacy sensitive ones) are locked. Neither long, semi-random alphanumeric passwords (which would be highly secure but not usable for mobile devices) nor swipe-only lockscreens (usable, but not offering any security) are advisable. Therefore, it is critically important for the lockscreen to strike a reasonable balance between security and usability, as it enables further authentication on higher levels.

\subsubsection{Tiered lockscreen authentication}
Towards this end, recent Android releases use a tiered authentication model where a secure knowledge-factor based authentication mechanism can be backed by convenience modalities that are functionally constrained based on the level of security they provide. The added convenience afforded by such a model helps drive lockscreen adoption and allows more users to benefit both from the immediate security benefits of a lockscreen and from features such as file-based encryption that rely on the presence of an underlying user-supplied credential. As of August 2020, starting with Android~7 we see that 77\% of devices with fingerprint sensors have a secure lockscreen enabled, while only 54\% of devices without fingerprints have a secure lockscreen\footnote{These numbers are from internal analysis that has not yet been formally published.}. 

As of Android~10, the tiered authentication model splits modalities into three tiers.
\begin{itemize}
	\item \emph{Primary Authentication} modalities are restricted to knowledge-factors and by default include password, PIN, and pattern\footnote{We explicitly refer to patterns connecting multiple dots in a matrix, not the whole-screen swipe-only lockscreen interaction that does not offer any security.}. Primary authentication provides access to all functions on the phone. It is well-known that the security/usability-balance of these variants is different: complex passwords have the highest entropy but worst usability, while PINs and patterns are a middle balance but may suffer e.g.\ from smudge~\cite{10.5555/1925004.1925009} ([T.P2]--[T.P3]) or shoulder surfing attacks~\cite{10.1145/3173574.3173738,10.1145/3025453.3025636} ([T.P1]). However, a knowledge-factor is still considered a trust anchor for device security and therefore the only one able to unlock a device from a previously fully locked state (e.g.\ from being powered off).
	
	\item \emph{Secondary Authentication} modalities are biometrics --- which offer easier, but potentially less secure (than Primary Authentication), access into a user's device\footnote{While the entropy of short passwords or PINs may be comparable to or even lower than for good biometric modalities and spoofability based on previous recordings is a potential issue for both, knowledge factors used as primary authentication offer two specific advantages: a) knowledge factors can be changed either (semi-) regularly or after a compromise has become known, but biometrics can typically not --- hence biometric identification is not generally considered a secret; b) knowledge factors support trivial, bit-for-bit comparison in simple code and hardware (cf.\ use of TRH as described in Section~\ref{subsec:hardware-sandbox}) instead of complex machine learning methods for state-of-the-art biometric sensors with liveness detection --- this simplicity leaves less room for implementation errors and other attack surface. Additionally, this perfect recall of knowledge factors allows cryptographic key material, e.g.\ for file encryption, to be directly entangled respectively derived from them.}. Secondary modalities are themselves split into sub-tiers based on how secure they are, as measured along two axes:
	\begin{itemize}
		\item \emph{Spoofability} as measured by the Spoof Acceptance Rate (SAR) of the modality~\cite{androidblogpost-biometrics-2018}. Accounting for an explicit attacker in the threat model on the level of [T.P2]--[T.P3] helps reduce the potential for insecure unlock methods~\cite{mayrhofer2020adversary}.
		\item \emph{Security of the biometric pipeline}, where a biometric pipeline is considered secure if neither platform or kernel compromise confers the ability to read raw biometric data or inject data into the biometric pipeline to influence an authentication decision. 
	\end{itemize}
	
	These axes are used to categorize secondary authentication modalities into three sub-tiers, where each sub-tier has constraints applied in proportion to their level of security~\cite{androidblogpost-biometrics-2020}:
	\begin{itemize}
		\item Class 3 (formerly `strong'): SAR<7\% and secure pipeline
		\item Class 2 (formerly `weak'): 7\%<SAR<20\% and secure pipeline
		\item Class 1 (formerly `convenience'): SAR>20\% or insecure pipeline
	\end{itemize}
	All classes are required to have a (na\"ive/random) false acceptance rate (FAR) of at most 1/50000 and a false rejection rate (FRR) of less than 10\%. Biometric modalities not meeting these minimum requirements cannot be used as Android unlock methods.
	Secondary modalities are prevented from performing some actions --- for example, they cannot decrypt file-based or full-disk encrypted user data partitions (such as on first boot) and are required to fallback to primary authentication once every 72 (Class~3) or 24 (Class~1 and~2) hours. Only Class~3 biometrics can unlock Keymint auth-bound keys and only Class~3 and~2 can be used for in-app authentication.

    The benefits of using a secondary modality in addition to a primary modality aren't limited to convenience: they help mitigate against the aforementioned smudge and shoulder surfing attacks by reducing the number of opportunities for an attacker to capture primary authentication attempts.
	
	Android 10 introduced support for implicit biometric modalities in \texttt{BiometricPrompt} for modalities that do not require explicit interaction, for example face recognition.
	Android 11 further introduced new features such as allowing developers to specify the authentication types accepted by their apps and thus the preferred level of security~\cite{url-android-biometric-auth-11}.
	
	\item \emph{Tertiary Authentication} modalities are alternate modalities such as unlocking when paired with a trusted Bluetooth device, or unlocking at trusted locations; they are also referred to as environmental authentication. Tertiary modalities are subject to all the constraints of secondary modalities. Additionally, like the weaker secondary modalities, tertiary modalities are also restricted from granting access to Keymint auth-bound keys (such as those required for payments) and also require a fallback to primary authentication after any 4-hour idle period. Android~10 switched tertiary authentication from an active unlock mechanism into an extending unlock mechanism that can only keep a device unlocked for a longer duration (up to 4 hours) but no longer unlock it once it has been locked.
\end{itemize}

The Android lockscreen is currently implemented by Android system components above the kernel, specifically \texttt{Keyguard} and the respective unlock methods (some of which may be OEM specific). User knowledge factors of secure lockscreens are passed on to Gatekeeper/Weaver (explained below in Section~\ref{subsec:hardware-sandbox}) both for matching them with stored templates and for deriving keys for storage encryption. One implication is that a kernel compromise could lead to bypassing the lockscreen --- but only after the user has logged in for the first time after reboot. 

\paragraph{Android devices as a second factor}
As of April 2019, lockscreen authentication on Android 7+ can be used for FIDO2/WebAuthn \cite{fidoalliance-android-certified,webauthn} authentication to web pages, additionally making Android phones second authentication factors for desktop browsers through implementing the Client to Authenticator Protocol (CTAP) ~\cite{googlecloudeblogpost-phone-as-a-key}. While this support is currently implemented in Google Play Services~\cite{url-fido-api}, the intention is to include support directly in AOSP in the future when standards have sufficiently settled down to become stable for the release cycle of multiple Android releases.

\paragraph{Identity Credential}
While the lockscreen is the primary means for user-to-device (U2D) authentication and various methods support device-to-device (D2D) authentication (both between clients and client/server authentication such as through WebAuthn), identifying the device owner to other parties has not been in focus so far. Through the release of a JetPack library\footnote{Available at \url{https://developer.android.com/jetpack/androidx/releases/security}}, apps can make use of a new ``Identity Credential'' subsystem to support privacy-first identification~\cite{Hoelzl2017IFIPSCPost} (and, to a certain degree, authentication). One example are third-party apps to support mobile driving licenses (mDL) according to the ISO 18013-5 standard~\cite{std-mdl}. The first version of this subsystem targets in-person presentation of credentials, and identification to automated verification systems is subject to future work.

Android~11 started including the Identity Credential subsystem in the form of a new HAL, a new system daemon, and API support in AOSP~\cite{androidblogpost-mobile-driving-license,url-identity-credentials-api}. If the hardware supports direct connections between the NFC controller and tamper-resistant dedicated hardware, credentials will be able to be marked for ``Direct Access''\footnote{See the HAL definition at \url{https://android-review.googlesource.com/c/platform/hardware/interfaces/+/1151485/30/identity/1.0/IIdentityCredentialStore.hal}.} to be available even when the main application processor is no longer powered (e.g.\ in a low-battery case).

In Android~13, an API for presenting multiple documents in a single session was added for improved usability in more complex scenarios (e.g.\ entrance tickets presented together with a nationally issued photo ID), and Android~14 adds ECDSA authentication in addition to the previously standardized MAC authentication for the underlying \texttt{mdoc} structures.

\subsection{Isolation and Containment}
\label{subec:isolation-and-containment}
One of the most important parts of enforcing the security model is to enforce it at runtime against potentially malicious code already running on the device. The Linux kernel provides much of the foundation and structure upon which Android's security model is based. Process isolation, and specifically of processes with different UIDs, provides the fundamental security primitive for sandboxing. With very few exceptions, the UID/process boundary is where security decisions are made and enforced --- Android intentionally does not rely on in-process compartmentalization such as the Java security model. The security boundary of a process is comprised of the process boundary and its entry points and implements \hyperref[rule:5]{rule \circled{5} (apps as security principals)} and \hyperref[rule:2]{rule \circled{2} (open ecosystem)}: an app does not have to be vetted or pre-processed to run within the sandbox, because it is contained within its own UID and associated process isolation.
Strengthening this boundary can be achieved by a number of means such as:
\begin{itemize}
	\item Access control: adding permission checks, increasing the granularity of permission checks, or switching to safer defaults (e.g.\ default deny) to address the full range of threats [T.A1]--[T.A7] and [T.D1]--[T.D2].
	\item Attack surface reduction: reducing the number of entry points, particularly [T.A1], [T.A2], [T.A7], and [T.A8].
	\item Containment: isolating and de-privileging components, particularly ones that handle untrusted content as in [T.A3] and [T.D2].
	\item Architectural decomposition: breaking privileged processes into less privileged components and applying attack surface reduction for [T.A2]--[T.A8] and [T.D2], i.e.\ the principle of least privilege.
	\item Separation of concerns: avoiding duplication of functionality.
\end{itemize}

\begin{figure}
	\centering
	\includegraphics[width=0.95\textwidth]{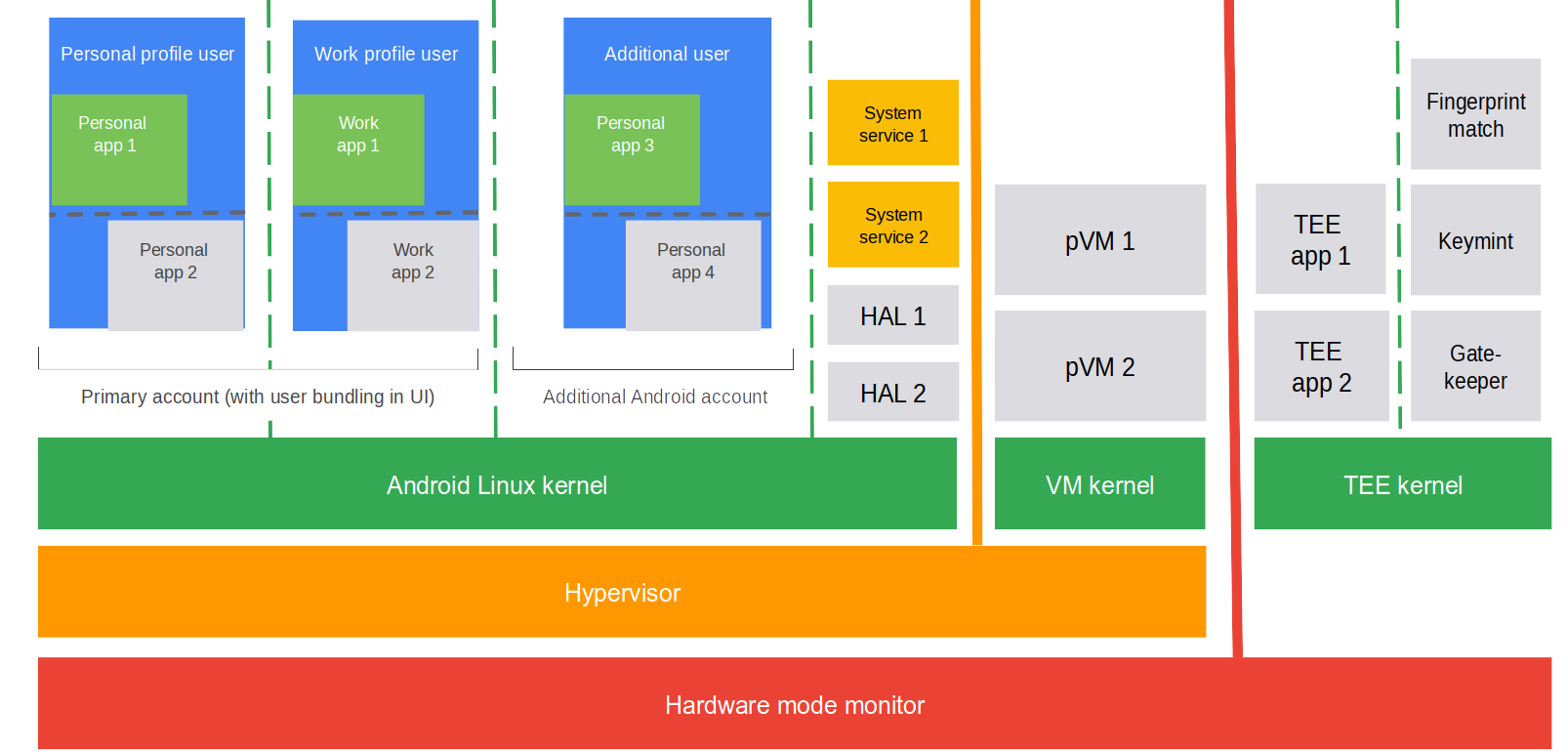}
	\Description[Sandboxing layers]{Layers of sandboxing}
	\caption{Layers of sandboxing}
	\label{fig:sandboxing-layers}
\end{figure}

In this section we describe the various sandboxing and access control mechanisms used on Android on different layers and how they improve the overall security posture. Figure~\ref{fig:sandboxing-layers} summarizes the multiple layers of sandboxing above and below the Linux kernel.

\subsubsection{Access control}
\label{subsec:permissions}
Android uses three distinct permission mechanisms to perform access control:
\begin{itemize}
	\item \textbf{Discretionary Access Control (DAC):}
	Apps run within their associated UNIX user ID (UID). App processes may grant or deny access to resources that are owned by their UID, by modifying permissions on the object (e.g., granting world read access) or by passing a handle to the object over IPC. On Android this is implemented using UNIX-style permissions that are enforced by the kernel. Processes running as the \texttt{root} user often have broad authority to override UNIX permissions of any other UID (subject to MAC permissions).
	\item \textbf{Mandatory Access Control (MAC):}
	The system has a security policy that dictates what actions are allowed. Only actions explicitly granted by policy are allowed. On Android this is implemented using SELinux~\cite{smalley_seandroid} and primarily enforced by the kernel. Android makes extensive use of SELinux to protect system components and assert security model requirements during compatibility testing. SELinux based MAC supports enforcing security boundaries even between processes running with the same UID, and therefore offers finer granularity of isolation.
	\item \textbf{Android permissions}
	add higher-level semantic permissions (such as location or camera access) that are granted to UIDs, as well as URI permission grants to provide the core mechanism for fine-grained access control, allowing an app to grant selective access to pieces of data it controls. Enforcement is primarily done in userspace by the data/service provider (with notable exceptions such as \texttt{INTERNET}). Permissions are defined statically in an app's \texttt{AndroidManifest.xml}~\cite{url-android-manifest}. As Android permissions are the most direct user-visible component of the access control elements, we describe them in more detail in the following Section~\ref{subsec:android-permissions}.
\end{itemize}

Each of the three permission mechanisms roughly aligns with one of the three parties of the multi-party authorization (\hyperref[rule:1]{rule \circled{1}}). The platform utilizes MAC, apps use DAC, and users authorize by granting Android permissions. Note that permissions are not intended to be a complete mechanism for obtaining consent in the legal sense, but a technical measure to enforce auditability and control. It is up to the app developer processing personal user data to meet applicable legal requirements.

\subsubsection{Android permissions}
\label{subsec:android-permissions}
In the early development of smartphones, protection from apps was largely seen as a security issue. The two common approaches being used were requiring that apps be signed by an entity that is vouching for the safety of those apps or prompting the user at runtime for the capabilities that the app needs. Since Android is an open platform that allows native code, relying on protection from an external entity was not viable; at the same time, implementations that prompted for every potentially unsafe operation may lead to a large number of confusing prompts to the user and quick user prompt-fatigue.

Android initially took an alternative approach of using install-time permissions. In this model, the user is shown the capabilities that an application will be granted prior to installing it, so they can look holistically at what the app will be able to do on their device and decide if that makes sense for what it claims to be. This worked fairly well in the initial mobile world of small, targeted apps: these apps tended to do a very focused thing (such as a contact manager, a game, a music player, etc.) and would stand out if they needed other capabilities that didn't match their functionality.

Over time, two changes happened: First, mobile applications became increasingly complex with a growing number of secondary features, leading to increasing numbers of install-time permissions for features a particular person may never use, and for which they were not sufficiently equipped to make such a decision at installation time (cf.~\cite{Felt:2012:APU:2335356.2335360,Felt2012HowTA,Roesner_user-drivenaccess,190982}). Second, privacy became a growing concern for operating systems, resulting in a smaller set of capabilities revolving around the user's personal information that are of particular interest.

Privacy changes the user experience with applications from ``Is this app safe for me to install?'' to ``I want to decide what of my personal information this app gets access to based on what I am doing with it.''  As a result of these changes in application behavior and user needs, Android~6.0 re-arranged its existing raw permissions from a security-oriented organization to privacy-oriented. This involved two major changes: (1) demoting pure security permissions to be invisible to the user and used only for auditing of applications; and (2) identifying a small set of types of clearly identifiable user information (such as location, contacts, camera) that the remaining permissions can be organized under and introducing runtime prompts the application must perform when needed for the user to allow that access.

A further refinement in Android permissions was introduced with Android~10 in the form of non-binary, context dependent permissions: in addition to \emph{Allow} and \emph{Deny}, some permissions (particularly location, and starting with Android~11 others like camera and microphone) can now be set to \emph{Allow only while using the app}. This third state only grants the permission when an app is in the foreground, i.e.\ when it either has a visible activity or runs a foreground service with permanent notification~\cite{androidblogpost-location-privacy-2019}. Android~11 extended this direction with one-time permissions that are granted until the app loses its foreground state.
	
\medskip
	
At a high level Android permissions fall into one of five classes in increasing order of severity, whose availability is defined by their \texttt{protectionLevel} attribute~\cite{url-android-manifest-permissions} with two parts (the protection level itself and a number of optional flags):
\begin{enumerate}
    \item \emph{Audit-only permissions}: These are install time permissions with protection level \texttt{normal} that are mostly related to security boundaries for apps and less relevant to user privacy decision, and which are thus granted automatically at install time. They are primarily used for auditability of app behavior.
		
    \item \emph{Runtime permissions}: These are permissions with protection level \texttt{dangerous} and apps must both declare them in their manifest as well as request users grant them during use. These permissions are guarding commonly used sensitive user data, and depending on how critical they are for the current functioning of an application, different strategies for requesting them are recommended~\cite{url-permission-request-guidance}. While runtime permissions are fairly fine-grained to support auditing and enforcement in-depth, they are grouped into logical permissions using the \texttt{permissionGroup} attribute. When requesting runtime permissions, the group appears as a single permission to avoid over-prompting.
		
    \item \emph{Special permissions}: For resources that are either considered higher risk than those protected by runtime permissions or that are otherwise special cases, there exists a separate class of permissions with much higher granting friction than the permission dialogs for runtime permissions. %
    In order for a user to allow an application to use a special access permission, the user must go to settings and manually grant the permission to the application. Specific examples to special access permissions are device admin, notification listeners, unrestricted network data access, or---critically---installing other packages. While the particular special permissions differ significantly, a commonality is that they are either niche use cases and/or too complicated to present as a runtime permission in the sense that they are not gating access to some clear pieces of personal data that users can directly relate to.

    \item \emph{Privileged permissions}: These permissions are for pre-installed applications only and allow privileged actions such as modifying secure settings or carrier billing.
        They typically cannot be granted by users during run-time but OEMs grant them by allowlisting the \texttt{privileged} permissions for individual apps~\cite{url-android-priv-perm-whitelist} in the system image.
        
        \texttt{Privileged} protection level permissions are usually coupled with the \texttt{signature} level.
		
    \item \emph{Signature permissions}: These permissions with protection level \texttt{signature} are only available to components signed with the same key as the component which declares the permission (i.e., the platform or an application) --- which is the platform signing key for platform permissions. They are intended to guard internal or highly privileged actions (e.g.\ configuring the network interfaces) and are granted at install time if the application is allowed to use them.
\end{enumerate}
	
Additionally, there are a number of protection flags that modify the grantability of permissions. For example, the \texttt{BLUETOOTH\_PRIVILEGED} permission has a \texttt{protectionLevel} of ``\texttt{signature or privileged}'', with the \texttt{privileged} flag allowing privileged applications to be granted the permission (even if they are not signed with the platform key).

\subsubsection{Application sandbox}
\label{subsec:app-sandbox}
Android's original DAC application sandbox separated apps from each other and the system by providing each application with a unique UID and a directory owned by the app. This approach was quite different from the traditional desktop approach of running applications using the UID of the physical user. The unique per-app UID simplifies permission checking and eliminates per-process ID (PID) checks, which are often prone to race conditions. Permissions granted to an app are stored in a centralized location (\texttt{/data/system/packages.xml}) to be queried by other services. For example, when an app requests location from the location service, the location service queries the permissions service to see if the requesting UID has been granted the location permission.

Starting with Android~4, UIDs are also used for separating multiple physical device users. As the Linux kernel only supports a single numeric range for UID values, device users and profiles %
are separated through a larger offset (\texttt{AID\_USER\_OFFSET=100000} as defined in AOSP source\footnote{See \texttt{system/core/include/private/android\_filesystem\_config.h} in the AOSP source tree.}) and apps installed for each user are assigned UIDs in a defined range (from \texttt{AID\_APP\_START=10000} to \texttt{AID\_APP\_END=19999}) relative to the device user offset. This combination is referred to as the Android ID (AID).

\paragraph{Users and profiles}
Android allows device sharing through the concept of separate users\footnote{The term ``user'' is somewhat overloaded in Android. It can be used to either refer to the person that is using the device, to the app sandbox (implemented as a Linux user), or to user separation on the device as explained in this section. While the rest of this paper uses the term to refer to the person using the device, this section uses it to refer to user separation on the device, calling the person the ``device user''.}, generally mapping to different people using the device. Users inherently do not change anything about the Android security model: the main security principal is the app sandbox, and access controls are enforced at that level. Users act as a logical grouping of app sandboxes\footnote{In fact, this is exactly how user separation is implemented in Android: ``users'' are defined by ranges of UIDs for their apps.}, guarded by a separate lock screen (and corresponding key), and treated appropriately by system services. Generally, the system enforces a few additional security properties on top of the regular app sandbox for apps in separate users:

\begin{itemize}
    \item Different users are separated in the system UI. Apps running on behalf of one user are not able to display alongside apps running on behalf of another user.
    \item Apps running in different users act as if they run on separate devices: communication between apps running in different users is not possible, and apps running in different users are not visible to each other.
\end{itemize}

A \emph{profile}, most commonly known in the form of a work profile, is a special case of a user, intended for separation of different \emph{personas} belonging to the same person. Contrary to normal users, profiles are presented directly alongside each other. They share the same system UI, including the launcher, the notification stream, and most device preferences. A secondary profile can optionally be accessed directly after unlocking the primary profile, while still having separate storage encryption keys. Interaction between different profiles is always mediated by the device user:

\begin{itemize}
    \item Direct communication between two instances of the same app running in separate profiles is gated by user opt-in.
    \item Direct communication between different apps running in separate profiles is not possible.
    \item The device user can explicitly and deliberately share data between profiles by using the share sheet.
\end{itemize}

Because of these strong data separation guarantees, enterprise policies are applied only on a per-profile (per-user) basis. The only exceptions to this are cases that improve the security of the device as a whole, such as defining a maximum screen timeout or network security requirements.

\paragraph{Historical development}
The UID/AID sandbox laid the groundwork and is still the primary enforcement mechanism that separates apps from each other. It has proven to be a solid foundation upon which to add additional sandbox restrictions. However, there are some limitations based on the traditional UNIX UID model: Processes running as root were essentially unsandboxed and possessed extensive power to manipulate the system, apps, and private app data. Likewise, processes running as the system UID were exempt from Android permission checks and permitted to perform many privileged operations. The use of DAC meant that apps and system processes could override safe defaults and were more susceptible to dangerous behavior, such as symlink following or leaking files/data across security boundaries via IPC or \texttt{fork}/\texttt{exec}. Additionally, DAC mechanisms can only apply to files on file systems that support access controls lists (respectively simple UNIX access bits). The main implication is that the FAT family of file systems, which is still commonly used on extended storage such as (micro-)~SD cards or media connected through USB, does not directly support applying DAC. On Android, each app has a well-known directory on external storage devices, where the package name of the app is included into the path (e.g.\ \texttt{/sdcard/Android/data/com.example}). Since the OS already maintains a mapping from package name to UID, it can assign UID ownership to all files in these well-known directories, effectively creating a DAC on a filesystem that doesn't natively support it. From Android~4.4 to Android~7, this mapping was implemented through FUSE, while Android~8.0 and later implement an in-kernel \texttt{sdcardfs} for better performance. Both are equivalent in maintaining the mapping of app UIDs to implement effective DAC. Android~10 introduced \emph{scoped storage}, which further limits access to external storage by allowing apps to access only their own external directory path and the media files created by themselves in the shared media store. %

\begin{table*}[ht]
\renewcommand{\arraystretch}{1.1} %

\begin{tabularx}{\linewidth}{cXl}
    \bf Release & \multicolumn{1}{c}{\bf Improvement}  &  \parbox[c]{1.4cm}{\centering\bf Threats \\ Mitigated}  \\ 
    \toprule

    1.0 & Core app sandboxing model defined, based on kernel UID separation & \parbox[t]{2.0cm}{[T.A2][T.A5] [T.A6][T.A7]} \\

    4.1 & Isolated process~\cite{android-isolated-process}: Apps may run services in a process with no Android permissions and access to only two binder services.
    For example, the Chrome browser runs its renderer in an isolated process for untrusted web content. & \parbox[t]{2.0cm}{[T.A3] access to [T.N1]  [T.A2][T.A5] [T.A6][T.A7]}\\ 
    
    5.x & SELinux enabled for all userspace processes, significantly improving the separation between apps and system processes. Boundaries between apps are still primarily enforced via UID sandbox. This also increased the auditability of policy, supporting analysis of security requirements during compatibility testing. & [T.A7][T.D2] \\ 
    
    6.x & SELinux restrictions on \texttt{ioctl}: 59\% of all app reachable kernel vulnerabilities were through the ioctl() syscall, and these restrictions limited the reachability of kernel vulnerabilities from user space code~\cite{talk-jeffv-lss-2016,talk-jeffv-lss-2015}. & [T.A7][T.D2] \\ 
    
    6.x & Removal of app access to \texttt{debugfs} (9\% of all app-reachable kernel vulnerabilities). & [T.A7][T.D2] \\ 

    7.x & \texttt{hidepid=2}: Remove \texttt{/proc/<pid>} side channel used to infer when apps were started. & [T.A4] \\ 
    
    7.x & perf-event-hardening (11\% of app reachable kernel vulnerabilities were reached via \texttt{perf\_event\_open()}). & [T.A7] \\

    8.x & All apps run with a \texttt{seccomp} filter, reducing kernel attack surface. & [T.A7][T.D2] \\ 
    
    9.0 & Per-app SELinux sandbox (for apps with \texttt{targetSdkVersion=P} or greater). & [T.A2][T.A4] \\ 
    
\bottomrule

\end{tabularx}
\caption{General Application sandboxing improvements in Android releases}\label{tab:application-sandboxing-improvements}
\end{table*}

\begin{table*}[ht]
\renewcommand{\arraystretch}{1.2} %

\begin{tabularx}{\linewidth}{cXl}
    \bf Release & \multicolumn{1}{c}{\bf Improvement}  &  \bf Threats Mitigated  \\ 
    \toprule

    5.x & Webview moved to an updatable APK, independent of a full system update. & [T.A3] \\ 
    
    6.x & Runtime permissions were introduced, which moved the request for dangerous permissions from install to first use. & [T.A1] \\ 
    
    6.x & Multi-user support: SELinux categories were introduced for a per-physical-user app sandbox.%
    & [T.P4] \\ 
    
    6.x & Safer defaults on private app data: App home directory moved from readable by all users, to only the app user (\texttt{0751} UNIX permissions to \texttt{0700}). & [T.A2] \\ 
    
    6.x & Moving SYSTEM\_ALERT\_WINDOW, WRITE\_SETTINGS, and CHANGE\_NETWORK\_STATE to special permission category. & [T.A1][T.A4] \\

    7.x & OPA/MITM CA certificates are not trusted by default. & [T.N2] \\ 
    
    7.x & Safer defaults on \texttt{/proc} filesystem access. & [T.A1][T.A4] \\

    8.x & Safer defaults on \texttt{/sys} filesystem access. & [T.A1][T.A4] \\ 

    8.x & Webviews for all apps move into the isolated process. & [T.A3] \\
    8.x & Apps must opt-in to use cleartext network traffic. & [T.N1] \\ 
    
    10 & Apps can only start a new activity with a visible window, in the foreground activity, or if more specific exceptions apply~\cite{url-activity-background-starts}.  & \parbox[t]{1.5cm}{[T.A2][T.A3] [T.A4][T.A7]} \\ 
    
    10 & File access on external storage is scoped to app-owned files. & [T.A1][T.A2] \\
    
    10 & Reading clipboard data is only possible for the app that currently has input focus or is the default input method (e.g. keyboard) app. & [T.A5] \\
    
    10 & \texttt{/proc/net} limitations and other side channel mitigations. & [T.A1] \\ 
    
    11 & Legacy access of non-scoped external storage is no longer available. & [T.A1][T.A2] \\
    
    11-13 & Restricted access to the hardware MAC address \cite{mac-address-availability}. & [T.D1] \\

    12 & Official support for Rust in AOSP. & [T.A7][T.D2] \\
    
    12-13 & Restrictions on passthrough touches and occluding windows to prevent Tapjacking attacks. & [T.A6] \\

\bottomrule

\end{tabularx}
\caption{App sandboxing that improved permissions, authorization limitations, and other improvements}\label{tab:application-sandboxing-improvements-perm}
\end{table*}

The primary UID sandbox limitations have been mitigated in a number of ways over subsequent releases, especially through the addition of MAC policies with SELinux in enforcing mode starting with Android~5, but also including many other mechanisms such as %
attack surface reduction (cf.\ Tables~\ref{tab:application-sandboxing-improvements} and~\ref{tab:application-sandboxing-improvements-perm}). In addition to SELinux, \texttt{seccomp} filters complement the MAC policy on a different level of syscall granularity. While the Chrome app is currently the main user of fine-grained \texttt{seccomp} filters, others can also use them to internally minimize attack surface for their components.

Another particular example for the interplay between DAC and MAC policies and changes based on lessons learned are the more recent restrictions to \texttt{ioctl}, \texttt{/proc}, and \texttt{/sys} since Android~7. As described more generally in Section~\ref{subsec:authorization}, limiting access to such internal interfaces improves app compatibility between platform versions and supports easier internal refactoring. For these kernel interfaces, restricting access had another benefit towards user privacy: while few apps used these kernel interfaces for legitimate purposes that could not be fulfilled with existing Android APIs, they were also abused by other apps for side-channel attacks~\cite{chen2014peeking, spreitzer2018procharvester, 236300, tuncay2020see} on data not otherwise accessible through their lack of required Android permissions (e.g.\ network hardware MAC addresses). Restricting access to these interfaces to follow an allow- instead of block-list approach is therefore a logical development in line with the defense-in-depth principle.

Rooting, as defined above, has the main aim of enabling certain apps and their processes to break out of this application sandbox in the sense of granting ``root'' user privileges~\cite{paper-iwssi2011}, which override the DAC rules (but not automatically MAC policies, which led to extended rooting schemes with processes intentionally exempt from MAC restrictions). Malware may try to apply these rooting approaches through temporary or permanent exploits and therefore bypass the application sandbox. 

\subsubsection{Sandboxing app SDKs}
\label{subsec:sdk-sandbox}
Generally, libraries that are embedded by an app are considered to be within the app's security boundary by the platform. Indeed, they are part of the app's code, signed by the app developer. One exception to this is the ads SDK runtime, introduced as part of Android~13. The ads SDK runtime allows an app to load code from third-party libraries, called ``SDKs'', in a secondary bound application sandbox with a separate UID. These third-party libraries are distributed as separate APKs, listed as dependencies in the app's manifest, and loaded on request by the platform at runtime.

Note that this does not change Android's multi-party authorization model: it only introduces the SDK\footnote{Or, technically, the union of all SDKs for a single app, as we will explain later.} as its own security principal, further using the architectural decomposition and containment strategies described above. Indeed, this is not very different from running the SDK as its own separate app, bound to the main app, even if the SDKs are loaded by the platform on behalf of the app. However, the ads SDK runtime environment introduces its own set of restrictions for ads that are different from regular apps, including a separate SELinux policy that minimizes the ability for fingerprinting and profiling, and a very limited set of permissions (i.e., audit-only permissions, which are of the protection level \texttt{normal}). The platform also introduces restrictions on the ability for separate sandboxes to interact across apps.

To enable use cases where the ads SDK would need to render part of the app, the platform allows the app to designate an area within its own view hierarchy in which the SDK is allowed to remotely render content. The composition of that content is entirely done by the platform, meaning that the app has no visibility into (or control over) the SDK's \texttt{View}s, and vice versa.

With this setup, the platform and the app distribution mechanism can now reason about the app and its libraries as two separate entities. This additional security boundary brings with it some desirable properties to the app developer, to the SDK developers, and to users.
For app developers, the largest benefit stems from the fact that the libraries no longer have access to the memory and private storage of the app (including, for example, its authentication tokens). Due to the way the platform remotely renders content from the SDK into the app, the app developer can be certain that the SDK is not able to access any other app content displayed inside its own view hierarchy. It is also advantageous for the app developer to no longer be considered as the party that is responsible %
for the SDK's code, and that app stores can enforce policies for them as separate entities. %
As a side effect, the SDK runtime helps increase the apps' stability by ensuring that SDK crashes do not crash the corresponding app.

SDK developers benefit from the fact that their SDK can now be distributed, and thus updated, separately from the app (as long as the API contract remains intact). To a lesser extent, they now have some guarantees that code and memory is not susceptible to tampering by the app developer, which is especially important for advertising SDKs that want to prevent abuse (e.g. programmatic clicks).

Users have the benefit of being able to apply a separate set of privacy restrictions to the app and the SDKs. For example, they can decide to allow the app to access their location, without inadvertently giving the SDKs that capability as well. As a side effect, the user also benefits from decreased storage usage, as multiple apps depending on the same SDK can share a dependency on the same APK.

Note that there is no security boundary between SDKs \emph{for the same app}. This is a deliberate tradeoff: while separating every SDK into its own process would provide many additional security benefits (e.g. preventing different SDKs from inspecting each other's memory), it would also have an unreasonable memory overhead. Because of this, the SDK runtime implements a few in-process mitigations, designed to ensure that SDK developers don't accidentally handle each others' resources. These mitigations include a separate classloader, \texttt{Context} object, and storage location for each SDK, and restrictions on executing native code. These mitigations are intended to prevent developers from accidentally sharing resources, and to facilitate policy enforcement on a per-SDK basis. As mentioned before, Android intentionally does not rely on in-process compartmentalization to enforce security boundaries.

The platform restricts the ability for SDKs to communicate across different SDK runtimes, or for them to communicate with apps other than the one they're loaded by. Note that communication between the app and its SDKs is not restricted in any way, and is enabled through a generic communication channel. Generally, collusion between app and SDK developer is out of scope, as the app developer can just as easily embed the SDK directly, or share a unique ID with the SDK allowing them to communicate out-of-band.

\subsubsection{Sandboxing system processes}
\label{subsec:system-process-sandbox}
\begin{figure}
	\centering
	\includegraphics[width=0.7\textwidth]{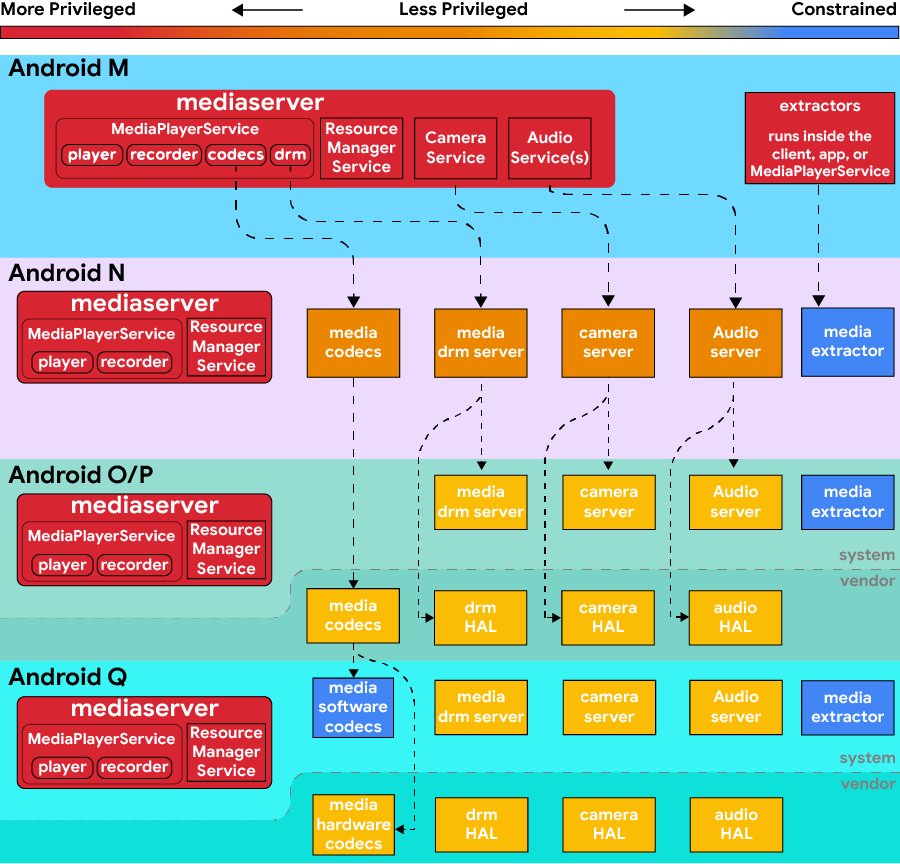}
	\Description[Summary of changes to mediaserver and codec sandboxing]{Changes to mediaserver and codec sandboxing from Android 6 to Android 10}
	\caption{Changes to mediaserver and codec sandboxing from Android 6 to Android 10}
	\label{fig:mediaserver-sandboxing}
\end{figure}

In addition to the application sandbox, Android launched with a limited set of UID sandboxes for system processes. Notably, Android's architects recognized the inherent risk of processing untrusted media content and so isolated the media frameworks into UID \texttt{AID\_MEDIA}, and this sandboxing has been strengthened from release to release with continuously more fine-grained isolation~\cite{androidblogpost-queue-hardening-2019}. Figure~\ref{fig:mediaserver-sandboxing} gives an overview of the sandboxing and isolation improvements for the media server and codecs. Other processes that warranted UID isolation include the telephony stack, Wi-Fi, and Bluetooth (cf.\ Table~\ref{tab:system-sandboxing-improvements}).

\begin{table*}[h!]
\renewcommand{\arraystretch}{1.2} %
\begin{tabularx}{\linewidth}{cXl}
    \bf Release & \multicolumn{1}{c}{\bf Improvement}  & \bf Threats Mitigated \\ 
\toprule
    4.4 & SELinux in enforcing mode: MAC for 4 root processes \texttt{installd}, \texttt{netd}, \texttt{vold}, \texttt{zygote}. & [T.A1][T.A7][T.D2]\\
    5.x & SELinux: MAC for all userspace processes. & [T.A1][T.A7] \\
    6.x & SELinux: MAC for all processes. & \\
    7.x & Architectural decomposition of mediaserver. & [T.A1][T.A7][T.D2] \\
    7.x & \texttt{ioctl} system call restrictions for system components~\cite{talk-jeffv-lss-2015}. & [T.A1][T.A7][T.D2] \\
    8.x & \emph{Treble} Architectural decomposition: Moved HALs (Hardware Abstraction Layer components) into separate processes, reduced permissions, restrict access to hardware drivers~\cite{talk-dcashman-lss-2017,androidblogpost-shut-the-hal-2017}. & [T.A1][T.A7][T.D2] \\
    10 & Software codecs (the source of approximately 80\% of the critical/high severity vulnerabilities in media components) were moved into a constrained sandbox. & [T.A7][T.D2] \\ 
    10 & Bounds Sanitizer (BoundSan): Missing or incorrect bounds checks on arrays accounted for 34\% of Android's userspace security vulnerabilities. Clang's BoundSan adds bounds checking on arrays when the size can be determined at compile time; enabled across the Bluetooth stack and in 11 software codecs. & [T.A7][T.D2] \\ 
    10 & Integer Overflow Sanitizer (IOSAN): The process of applying IOSAN to the media frameworks began in Android 7.0 and was completed in Android 10. & [T.A7][T.D2] \\ 
    10 & Scudo included as a dynamic heap allocator designed to be resilient against heap related vulnerabilities. & [T.A7][T.D2] \\
    10 & Shadow Call Stack (SCS, protecting the call graph backwards edge by protecting return addresses) enabled for Bluetooth. & [T.A7][T.D2] \\
    12 & Shadow Call Stack enabled for NFC. & [T.A7][T.D2] \\
    12 & Memory Tagging Extensions (MTE)~\cite{arm-mte} software support added & \parbox[t]{2.0cm}{[T.A5][T.A6] [T.A7][T.D2]} \\
    13 & Boundsan and IOSAN: Compiler-based sanitizers enabled in critical attack surface of the cellular baseband in some devices. & [T.A7][T.D2] \\
    14 & Branch Target Identification (BTI) enabled by default on supported hardware, providing forward-edge protection by preventing execution of instructions which are not intended branch targets. & [T.A7][T.D2] \\
    14 & Return Pointer Authentication (PAC-RET) enabled by default on supported hardware. This provides backward-edge protection by signing the return address stored on the stack. & [T.A7][T.D2] \\
\bottomrule 
\end{tabularx}
\caption{System sandboxing improvements in Android releases}\label{tab:system-sandboxing-improvements}
\end{table*}

\subsubsection{Sandboxing specific private data processing}
\label{subsec:app-pcc}
Android~12 introduced Private Compute Core (PCC) as an isolated environment to maintain separation from apps while enabling users and developers to maintain control over their data~\cite{article-pcc-arxiv.2209.10317}. 
Its main use case is to provide a safe environment for deriving less sensitive, potentially pseudonymous (e.g.\ $k$-anonymous) data from sensitive raw ambient (sensor captured) or operating system level data.

The whole PCC is sandboxed from Android system services and other apps through framework APIs provided by the AOSP base, and all data flowing into or out of PCC needs to go through these open components. Apps running within this PCC sandbox, such as the Google-proprietary ``Android System Intelligence'' for processing ambient and other data for on-device machine learning\footnote{Particular services as of Android 12 include Google ``Live Caption'', ``Now Playing'', ``Smart Reply'', and ``Screen attention'', among others.}, are prevented from directly accessing any other resources except through the provided PCC APIs. This includes network access; components running in PCC do not receive the \texttt{INTERNET} permission and can therefore not open any network sockets. Any communication to external services is required to go through another open source layer called ``Private Compute Services'' (PCS)\footnote{\url{https://github.com/google/private-compute-services/}}, which is distributed as an APK and includes standard support for federated learning and analytics, private information retrieval (PIR) using homomorphic encryption, and secure download of pre-trained ML models.

While basic data flow is restricted by Android platform permissions available to PCC apps, more granular control can be implemented by these applications themselves (e.g.\ for deleting stored data) and/or additional Android security/privacy controls such as camera, microphone, and other sensor toggles. All such specific PCC data flows are expected to conform to the multi-party authorization model. One example demonstrating this granular authorization is the ``Content Capture API''\footnote{\url{https://developer.android.com/reference/android/view/contentcapture/ContentCaptureManager}} introduced in Android 11: it respects developer authorization through the \texttt{FLAG\_SECURE} opt-out flag\footnote{\url{https://developer.android.com/reference/android/view/WindowManager.LayoutParams\#FLAG SECURE}} and user authorization through the PCC privacy settings implemented in AOSP.

For further details on high-level designs and example use cases of PCC, we refer to the separate whitepaper~\cite{article-pcc-arxiv.2209.10317}. PCC applications can benefit from additional confidentiality guarantees for data storage and processing in protected virtual machines (see Section~\ref{subsec:virtual-machines-sandbox}) even under the assumption of a compromised Android user space when the particular implementation uses such pVMs for sandboxing.

\subsubsection{Sandboxing the kernel}
\label{subsec:kernel-sandbox}
Security hardening efforts in Android userspace have increasingly made the kernel a more attractive target for privilege escalation attacks~\cite{talk-jeffv-lss-2016}. Hardware drivers provided by System on a Chip (SoC) vendors accounted for the vast majority of kernel vulnerabilities on Android in 2018~\cite{talk-jeffv-lss-2018}. Reducing app and system access to these drivers was described above, but kernel-level drivers cannot be sandboxed within the kernel themselves, as Linux still is a monolithic kernel (as opposed to microkernel approaches). For this reason and others, significant improvements have been made to mitigate exploitation of weaknesses in all code running within kernel mode, including the core Linux kernel components and vendor drivers (cf.\ Table~\ref{tab:kernel-sandboxing-improvements}).

\begin{table*}[h]
\renewcommand{\arraystretch}{1.2} %
\begin{tabularx}{\linewidth}{cXl}
    \bf Release & \multicolumn{1}{c}{\bf Improvement}  & \bf Threats Mitigated \\ 
\toprule
    5.x & Privileged eXecute Never (PXN)~\cite{url-pxn}: Disallow the kernel from executing code in userspace memory. Prevents \emph{return-to-user (ret2usr)} style attacks. & [T.A7][T.D2] \\
    6.x & Kernel threads moved into SELinux enforcing mode, limiting kernel access to userspace files. & [T.A7][T.D2] \\
    8.x & Privileged Access Never (PAN) and PAN emulation: Prevent the kernel from accessing any userspace memory without going through hardened \texttt{copy-*-user()} functions~\cite{androidblogpost-kernel-hardening-2017}. & [T.A7][T.D2] \\
    9.0 & Control Flow Integrity (CFI): Ensures that front-edge control flow stays within a precomputed graph of allowed function calls~\cite{androidblogpost-kernel-hardening-cfi-2018}. & [T.A7][T.D2] \\
    10 & Shadow Call Stack (SCS): Protects the backwards edge of the call graph by protecting return addresses~\cite{androidblogpost-kernel-hardening-scs-2019}. & [T.A7][T.D2] \\
    11 & Require latest long-term support (LTS) kernel with security updates and bug fixes~\cite{android-enterprise-whitepaper-2023}. & [T.A7][T.P1] \\
    12 & Bounds Sanitizer: The local-bounds part of Bounds Sanitizer is enabled by default in the Android Generic Kernel Image (GKI). This provides runtime detection of out of bounds accesses when the allocation size is determinable at compile time. & [T.A7][T.D2] \\
    14 & Kernel Control Flow Integrity (KCFI): Providing similar protection to CFI, KCFI has been enabled as part of the Android GKI.  & [T.A7][T.D2] \\
\bottomrule 
\end{tabularx}
\caption{Kernel sandboxing improvements in Android releases}\label{tab:kernel-sandboxing-improvements}
\end{table*}

\subsubsection{Sandboxing below the kernel}
\label{subsec:hardware-sandbox}
In addition to the kernel, the trusted computing base (TCB) on Android devices starts with the bootloader, which is typically split into multiple stages, and implicitly includes other components below the kernel, such as the trusted execution environment (TEE), hardware drivers, and userspace components \texttt{init}, \texttt{ueventd}, and \texttt{vold}~\cite{aosp-docs-severity-tcb}. It is clear that the sum of all these creates sufficient complexity that, given current state of the art, we have to assume bugs in some of them. For highly sensitive use cases, even the mitigations against kernel and system process bugs described above may not provide sufficient assurance against potential vulnerabilities. 

Therefore, we explicitly consider the possibility of a kernel or other TCB component failure as part of the threat model for some select scenarios. Such failures explicitly include \emph{compromise} e.g.\ through directly attacking some kernel interfaces based on physical access in [T.P1], [T.P3], and [T.P4] or chaining together multiple bugs from user space code to reach kernel surfaces in [T.A7]; \emph{misconfiguration} e.g.\ with incorrect or overly permissive SELinux policies~\cite{Chen:2017:ASP:3134600.3134638}; or \emph{bypass} e.g.\ by modifying the boot chain to boot a different kernel with deactivated security policies. To be clear, with a compromised kernel or other TCB parts, Android no longer meets the compatibility requirements and many of the security and privacy assurances for users and apps no longer hold. However, we can still defend against some threats even under this assumption:

\begin{itemize}
	\item \textbf{Keymint} (née Keymaster) implements the Android keystore in TEE to guard cryptographic key storage and use in the case of a run-time kernel compromise~\cite{url-android-keystore}. That is, even with a fully compromised kernel, an attacker cannot read key material stored in Keymint\footnote{Note: This assumes that hardware itself is still trustworthy. Side-channel attacks such as \cite{10.1007/978-3-030-10970-7_11} are currently out of scope of this (software) platform security model, but influence some design decisions on the system level, e.g.\ to favor dedicated TRH over on-chip security partitioning.}. Apps can explicitly request keys to be stored in Keymint, i.e.\ to be hardware-bound, to be only accessible after user authentication (which is tied to Gatekeeper/Weaver), and/or request attestation certificates to verify these key properties~\cite{url-android-keystore-attestation}, allowing verification of compatibility in terms of \hyperref[rule:3]{rule \circled{3} (compatibility)}.
	
	\item \textbf{Strongbox}, specified starting with Android 9.0, implements the Android keystore in separate tamper resistant hardware (TRH) for even better isolation. This mitigates [T.P2] and [T.P3] against strong adversaries, e.g.\ against cold boot memory attacks~\cite{Halderman:2009:LWR:1506409.1506429} or hardware bugs such as Spectre/Meltdown~\cite{Lipp2018meltdown,Kocher2018spectre}, Rowhammer~\cite{van_der_veen_drammer:_2016,talk-rowhammer-trustzone}, or Clkscrew~\cite{clkscrew} that allow privilege escalation even from kernel to TEE. From a hardware perspective, the main application processor (AP) will always have a significantly larger attack surface than dedicated secure co-processor. Adding a separate TRH affords another sandboxing layer of defense in depth. 
	
	The Google Pixel~3 was the first device to support Strongbox with a dedicated TRH (Titan~M \cite{googleblogpost-titan-m-pixel-3}), and other OEM devices have since started to implement it, often using standard secure elements that have been available on Android devices for NFC payment and other use cases.
	
	\begin{leftbar}
		Note that only storing and using keys in TEE or TRH does not completely solve the problem of making them unusable under the assumption of a kernel compromise: if an attacker gains access to the low-level interfaces for communicating directly with Keymint or Strongbox, they can use it as an oracle for cryptographic operations that require the private key. This is the reason why keys can be authentication bound and/or require user presence verification, e.g.\ by pushing a hardware button that is detectable by the TRH to assure that keys are not used in the background without user authorization.
	\end{leftbar}
	
	\item \textbf{Gatekeeper} implements verification of user lock screen factors (PIN/password/pattern) in TEE and, upon successful authentication, communicates this to Keymint for releasing access to authentication bound keys~\cite{url-android-gatekeeper}. \textbf{Weaver} implements the same functionality in TRH and communicates with Strongbox. Specified for Android 9.0 and initially implemented on the Google Pixel~2 and newer phones, we also add a property called \emph{Insider Attack Resistance} (IAR): without knowledge of the user's lock screen factor, an upgrade to the Weaver/Strongbox code running in TRH will wipe the secrets used for on-device encryption~\cite{androidblogpost-insider-attack-resistance-2018,usenix-enigma2019-android-insider-attack-resistance}. That is, even with access to internal code signing keys, existing data cannot be exfiltrated without the user's cooperation, directly addressing threat [T.A8].
	
	\item \textbf{Protected Confirmation}, also introduced with Android 9.0~\cite{url-android-protected-confirmation}, partially mitigates [T.A4] and [T.A6]. In its current scope, apps can tie usage of a key stored in Keymint or Strongbox to the user confirming that they have seen a message displayed on the screen by pushing a physical button. Upon confirmation, the app receives a hash of the displayed message, which can be used to remotely verify that a user has confirmed the message. By controlling the screen output through TEE when protected confirmation is requested by an app, even a full kernel compromise (without user cooperation) cannot lead to creating these signed confirmations.
\end{itemize}

\subsubsection{Sandboxing other kernels}
\label{subsec:virtual-machines-sandbox}
Android has mandated the existence of a TEE since version~6.0~\cite[Android 6.0, section 7.3.10.]{url-cdd} in order to provide hardware-assisted isolation that has a low attack surface and is not dependent on the Linux kernel to uphold its security assurances (cf. Section~\ref{subsec:hardware-sandbox}). Unfortunately, as more and more functionality has moved into the TEE, its attack surface has increased over time, partially undermining the reason for its existence. A particular concern in terms of attack surface is that code running in the TEE typically has \emph{access to all physical RAM}, including the Android ``normal world'' side, and thus vulnerabilities in any of the TEE kernel or apps result in significantly increased risk. 

While some hardware access capabilities still require the TEE in its current form, many TEE apps do not require this level of privilege and could be compartmentalized into other domains. Much like it has been done in userspace, ideally we could split the TEE into multiple lower-privilege TEEs, but that is not currently feasible due to hardware constraints in widely deployed CPUs\footnote{Hypervisor based separation in the ARM secure world, i.e. SEL2, is not yet widely available in the Android ecosystem.}.

The \emph{Android Virtualization Framework}~\cite{url-android-virtualization-framework} was introduced in Android~13 to take advantage of the separation capabilities that could be provided by a hypervisor outside the TEE in the ARM architecture. Not only is ARM EL2 already widely deployed across the Android ecosystem, but it also allows to provide multiple distinct isolation units, solving the scaling problem that we currently have with TEE-based solutions.

The AOSP reference hypervisor implementation, pKVM, is based on the Linux kernel KVM hypervisor, but with some important improvements. During early boot, the hypervisor code is split out of the kernel, which initially runs at EL1, and is installed into EL2. Despite being a part of the Linux kernel codebase, the component that is installed in EL2 is small\footnote{The open source pKVM hypervisor is smaller than all currently used TEE kernels including the AOSP reference implementation in the form of Trusty~\cite{url-trusty}.}, an intentional decision to harden the hypervisor security by limiting its complexity and attack surface. Unlike normal KVM, pKVM enforces separation between all VMs, including the host role which is filled by the Android VM. One major advantage of this design is that the hypervisor code running in EL2 minimizes hardware dependent driver code, as early hardware initialization is performed by the Linux kernel code before splitting control between EL2 and EL1, further reducing the hypervisor attack surface.

Note that pKVM in Android~14 delegates scheduling to the standard Android Linux kernel. Denial of service of virtual machines from a manipulated Android kernel point of view is therefore outside the threat model of AVF in its current implementation. Dynamic memory management, on the other hand, is a key piece of the security guarantees provided by pKVM, providing isolation without the need for physically contiguous carve-outs in physical RAM. To this end, pKVM supports atomic state changes for virtual machines, including donating pages from the host (Android) to guests (application VMs) and vice versa and explicit page sharing between host and guests or the TEE.
To minimize attack surface in the early initialization code running within each protected VM (pVM), the AVF reference implementation of the pVM firmware has been rewritten in Android~14; initially based on the \texttt{U-Boot} bootloader, the new firmware is written in Rust to benefit from memory safety guarantees~\cite{androidblogpost-bare-metal-rust}

The system code in Java, including system server and boot classpath, are normally compiled on the OEM's server and protected by Verified Boot with a signature.
Since Android~10, parts of the operating system can be updated independently as \emph{Modular System Components}\footnote{\url{https://source.android.com/docs/core/ota/modular-system}}, meaning that some of this compilation, including most Java components, need to be compiled on-device. In order to ensure the same security properties and prevent persistent attacks, this local compilation needs to happen in a secure environment.

Before AVF/pKVM, the compilation could only happen at early boot when the device had only run trusted code protected by Verified Boot. The output was then signed with a key restricted to early boot to ensure compilation could be skipped in subsequent boots. However, this compilation process still significantly slowed down the first boot after the Modular System update.

AVF/pKVM enables \emph{Isolated Compilation}. It allows the compilation to happen safely in the background (even if Android is significantly compromised) after the Modular System component has been staged. The isolated environment allows running only trusted code that accepts only trusted input and flags. The compiled artifacts are signed, together with the information of the compilation context (e.g., version, flags, etc.). This allows even the first post-update Android boot to skip the compilation step after checking these artifacts.

\subsubsection{Sandboxing firmware on other processors}
\label{subsec:other-processors-sandbox}
Securing the platform requires going beyond the confines of the Application Processor (AP). Android's defense-in-depth strategy, as well as the scope of this paper, also applies to the firmware running on bare-metal environments in the micro-controllers that perform various specialized tasks, such as security functions, image and video processing, and cellular communication. Separation of concerns and clear interfaces to communicate between the different processors in the form of HALs (specified in AIDL) are the primary means of sandboxing on this level. 

Many of the exploit mitigations described in Section~\ref{subsec:exploit-mitigation} are increasingly applied to non-AP processors as well. While most of these firmware code bases are chipset/ODM/OEM specific, the Android platform supports applying sandboxing and exploit mitigation methods initially developed for AP code on lower level firmware, e.g.\ through the systematic application of compiler-based sanitizers in connectivity firmware (Table~\ref{tab:system-sandboxing-improvements}), and with explicit support for memory-safe languages like Rust~\cite{androidblogpost-hardening-connectivity-firmware}.

\subsection{Encryption of data at rest}
\label{subec:encryption-at-rest}
A second element of enforcing the security model, particularly rules \hyperref[rule:1]{\circled{1} (multi-party authorization)} and \hyperref[rule:3]{\circled{3} (compatibility)}, is required when the main system kernel is not running or is bypassed (e.g.\ by reading directly from non-volatile storage).

Full Disk Encryption (FDE) uses a credential protected key to encrypt the entire user data partition. FDE was introduced in Android~5.0, and while effective against [T.P2], it had a number of shortcomings. Core device functionality, such as incoming calls, accessibility services, and alarms, were inaccessible until password entry\footnote{Note that very limited functionality included only an emergency dialer for outgoing emergency calls, as this did not depend on availability of the data partition.}. Multi-user support introduced in Android~6.0 still required the password of the primary user before disk access.

These shortcomings were mitigated by File Based Encryption (FBE), introduced in Android~7.0. On devices with TEE or TRH, all keys are derived within these secure environments, entangling the user knowledge factor with hardware-bound random numbers that are inaccessible to the Android kernel and components above. %
 FBE allows individual files to be tied to the credentials of different users, cryptographically protecting per-user data on shared devices [T.P4]. Devices with FBE also support a feature called \emph{Direct Boot} which enables access to emergency dialer, accessibility services, alarms, and receiving calls all before the user inputs their credentials. 

Android 10 introduced support for Adiantium~\cite{Crowley_Biggers_2018}, a new wide-block cipher mode based on AES, ChaCha, and Poly1305 to enable full device encryption without hardware AES acceleration support. While this does not change encryption of data at rest for devices with existing AES support, lower-end processors can now also encrypt all data without prohibitive performance impact. The significant implication is that all devices shipping originally with Android 10 are required to encrypt all data by default without any further exemptions, homogenizing the Android ecosystem in that aspect.

Note that encryption of data at rest helps significantly with enforcing \hyperref[rule:4]{rule~\circled{4} (safe reset)}, as effectively wiping user data only requires to delete the master key material, which is much quicker and not subject to the complexities of e.g.\ flash translation layer interactions.

\subsection{Encryption of data in transit}
\label{subec:encryption-in-transit}
Android assumes that all networks are hostile and could be injecting attacks or spying on traffic. In order to ensure that network level adversaries do not bypass app data protections, Android takes the stance that \emph{all} network traffic should be end-to-end encrypted. Link level encryption is insufficient. This primarily protects against [T.N1] and [T.N2]. However, this is not sufficient protection against [T.N3] as certain types of user traffic over cellular networks (cf.\ Section~\ref{subec:cellular-network-security}) are strictly only protected by link level encryption such as circuit-switched voice and Short Message Service (SMS).

In addition to ensuring that connections use encryption, Android focuses heavily on ensuring that the encryption is used correctly. While TLS options are secure by default, we have seen that it is easy for developers to incorrectly customize TLS in a way that leaves their traffic vulnerable to OPA/MITM~\cite{GIJABS12,Fahl:2012:WEM:2382196.2382205,Fahl:2013:RSD:2508859.2516655}. Table~\ref{tab:network-security-improvements} lists platform changes in terms of making network connections safe by default, which have led to significant improvements of TLS usage in apps~\cite{androidblogpost-tls-adoption-2019}.

\begin{table*}[h]
\renewcommand{\arraystretch}{1.2} %
\begin{tabularx}{\linewidth}{cXl}
    \bf Release & \multicolumn{1}{c}{\bf Improvement}  & \bf Threats Mitigated \\ 
    \toprule

    6.x & \texttt{usesCleartextTraffic} in manifest to prevent unintentional cleartext connections~\cite{googleblogpost-nogotofail-2014}. & [T.N1][T.N2] \\
    7.x & Network security config~\cite{url-android-security-config} to declaratively specify TLS and cleartext settings on a per-domain or app-wide basis to customize TLS connections. & [T.N1][T.N2] \\
    9.0 & DNS-over-TLS~\cite{androidblogpost-dns-over-tls-2018} to reduce sensitive data sent over cleartext and made apps opt-in to using cleartext traffic in their network security config. & [T.N1][T.N2] \\
    9.0 & TLS is the default for all connections~\cite{androidblogpost-tls-by-default-2018}. & [T.N1][T.N2] \\ 
    10 & MAC randomization enabled by default for client mode, SoftAP, and Wi-Fi Direct \cite{url-mac-randomization}. & [T.P1][T.N1] \\
    10 & TLS 1.3 support. & [T.N1][T.N2] \\ %
    12 & Option to disable the 2G radio\footnote{Requires Radio HAL v1.6} \cite{android12_2G_toggle}. & [T.N3] \\
    11-13 & DNS-over-HTTP/3~\cite{androidblogpost-DoH3-2022} introduced as part of Android 13, and added to older platform versions through a Modular System Update & [T.N1][T.N2] \\
    14 & Option to reject null-ciphered cellular connections\footnote{Requires Radio HAL v2.1}. & [T.N3]\\
   
   \bottomrule 
\end{tabularx}
\caption{Network sandboxing improvements in Android releases}\label{tab:network-security-improvements}

\end{table*}

\subsection{Exploit mitigation}
\label{subsec:exploit-mitigation}
A robust security system should assume that software vulnerabilities exist and actively defends against them. Historically, about 85\% of security vulnerabilities on Android result from unsafe memory access (cf.~\cite[slide 54]{talk-nnk-blackhat-2016}). While this section primarily describes mitigations against memory unsafety ([T.P1-P4], [T.N2], [T.A1-A3,A7], [T.D2]) we note that the best defense is the memory safety offered by languages such as Java, Kotlin, or Rust. Much of the Android framework is written in Java, effectively defending large swathes of the OS from entire categories of security bugs.

Android mandates the use of a number of mitigations including ASLR~\cite{Bhatkar:2003:AOE:1251353.1251361,Shacham:2004:EAR:1030083.1030124}, RWX memory restrictions (e.g.\ $W \oplus X$, cf.~\cite{Seshadri:2007:STH:1294261.1294294}), and buffer overflow protections, such as stack-protector for the stack and allocator protections for the heap. Similar protections are mandated for Android kernels~\cite{androidblogpost-kernel-hardening-2017}.

In addition to the mitigations listed above, Android is selectively enabling new mitigations, focusing first on code areas which are remotely reachable (e.g.\ the media frameworks~\cite{androidblogpost-hardening-media-2016}) or have a history of high severity security vulnerabilities (e.g.\ the kernel). Android also strongly recommends its ecosystem partners similar levels of hardening in over the air remotely reachable firmware running on other processors within the SoC that perform various specialized
tasks (e.g.\ cellular communications)~\cite{androidblogpost-hardening-connectivity-firmware}.

Android has pioneered the use of LLVM undefined behavior sanitizer (UBSAN) and other address sanitizers~\cite{180957} in production devices to protect against vulnerabilities in the media frameworks, kernel, and other security sensitive components.

Android is also rolling out Control Flow Integrity (CFI)~\cite{androidblogpost-kernel-hardening-cfi-2018} in the kernel and security sensitive userspace components including media, Bluetooth, Wi-Fi, NFC, and parsers~\cite{androidblogpost-compiler-hardening-2018} in a fine-grained variant as implemented by current LLVM~\cite{184459} that improves upon previous, coarse-grained approaches that have been shown to be ineffective~\cite{184481}. 
Starting with Android~10, the common Android kernel as well as parts of the Bluetooth stack can additionally be protected against backwards-edge exploitation through the use of Shadow Call Stack (SCS), again as implemented by current LLVM~\cite{androidblogpost-queue-hardening-2019} as the best trade-off between performance overhead and effectiveness~\cite{8835389}. Android~11 started auto-initializing memory in C/C++ code~\cite{androidblogpost-memory-autoinit-2020}.

Android~12 introduced software support for Memory Tagging Extensions (MTE)~\cite{arm-mte}, an ARM CPU hardware implementation~\cite{arxiv-mte-abs-1802-09517} of tagged memory. MTE marks each memory allocation/ deallocation with additional metadata (a ``tag'' to a memory location), which then can be associated with pointers that reference that memory location. At runtime, the CPU checks that the pointer and the metadata tags match on each load and store. Google Pixel~8, originally shipping with Android~14, is the first device with MTE hardware support enabled and exposed to users as an optional developer option.

To completely remove many of the memory unsafety bug classes at compile time, Android started to include support for system components written in Rust instead of C/C++ in 2021~\cite{androidblogpost-rust-2021}, and Android~12 already featured official support for Rust as platform system programming language. Android~13 was the first release in which more \emph{new} code was written in memory safe languages (Rust, Java, or Kotlin) than memory unsafe languages (C/C++). At the time of this writing, no memory safety vulnerability has been been discovered in Android Rust code~\cite{androidblogpost-rust-2022}, indicating a significant improvement in platform level mitigations for such bug classes.\footnote{Other (including logical) vulnerabilities often have lower impact and severity than memory safety vulnerabilities.}

These code and runtime safety mitigation methods work in tandem with isolation and containment mechanisms (cf.\ Tables~\ref{tab:application-sandboxing-improvements} to~\ref{tab:kernel-sandboxing-improvements} for added mitigations over time) to form many layers of defense; even if one layer fails, other mechanisms aim to prevent a successful exploitation chain. Mitigation mechanisms also help to uphold rules \hyperref[rule:2]{\circled{2} (open ecosystem)} and \hyperref[rule:3]{\circled{3} (compatibility)} without placing additional assumptions on which languages apps are written in.

\medskip

There are other types of exploits than apps directly trying to circumvent security controls of the platform or other apps: malicious apps can try to mislead users through deceptive UI tactics to either receive authorization grants against users' interests (including tapjacking~\cite{tapjacking}, the app equivalent of clickjacking~\cite{fratantonio17:cloakdagger}) ([T.A4]--[T.A6], [T.D1]), existing legitimate apps can be repackaged together with malicious code ([T.A1]--[T.A2]), or look-alike and similarly named apps could try to get users to install them instead of other well-known apps. Such user deception is not only a problem in the Android ecosystem but more generally of any UI-based interaction. As deception attacks tend to develop and change quickly, platform mitigations are often too slow to roll out, making dynamic blocking more effective. Within the Android ecosystem, mitigations against such kinds of exploits are therefore based on multiple mechanisms, notably submission-time checks on Google Play and on-device run-time checks with Google Play Protect. Nonetheless, platform security has adapted over time to make certain classes of UI deception exploits harder or impossible, e.g.\ through restricting SYSTEM\_ALERT\_WINDOW, background activity limitations, scoped external storage, or occlusion / touch passthrough prevention (cf.\ Table~\ref{tab:application-sandboxing-improvements}).

\subsection{System integrity}
\label{subsec:integrity}
Finally, system (sometimes also referred to as device) integrity is an important defense against attackers gaining a persistent foothold. AOSP has supported \emph{Verified Boot} using the Linux kernel \texttt{dm-verity} support since Android KitKat, providing strong integrity enforcement for the Trusted Computing Base (TCB) and system components to implement \hyperref[rule:4]{rule~\circled{4} (safe reset)}. 
Verified Boot~\cite{url-verified-boot} has been mandated since Android Nougat (with an exemption granted to devices which cannot perform AES crypto above 50MiB/sec. up to Android~8, but no exemptions starting with Android~9.0) and makes modifications to the boot chain detectable by verifying the boot, TEE, and additional vendor/OEM partitions, as well as performing on-access verification of blocks on the system partition~\cite{url-verified-boot-spec}. That is, attackers cannot permanently modify the TCB even after all previous layers of defense have failed, in order to achieve a successful kernel compromise. Note that this assumes the primary boot loader as root of trust to still be intact. As this is typically implemented in a ROM mask in sufficiently simple code, critical bugs at that stage are less likely.

Additionally, rollback protection with hardware support (counters stored in tamper-proof persistent storage, e.g.\ a separate TRH as used for Strongbox or enforced through RPMB as implemented in a combination of TEE and eMMC controller~\cite{7411305}) prevents attacks from flashing a properly signed but outdated system image that has known vulnerabilities and could be exploited. Finally, the Verified Boot state is included in key attestation certificates (provided by Keymint/Strongbox) in the \texttt{deviceLocked} and \texttt{verifiedBootState} fields, which can be verified by apps as well as passed onto backend services to remotely verify boot integrity~\cite{url-android-key-attestation} and to support \hyperref[rule:3]{rule \circled{3} (compatibility)}. 

Starting with Android~10, on devices supporting Android Verified Boot\footnote{AVB, the recommended default implementation for verifying the integrity of read-only partitions~\cite{url-verified-boot-spec}} version 2, the VBMeta struct digest (a top-level hash over all parts) is included in key attestation certificates to support firmware transparency. This is done by verifying that the digest matches that of a released firmware image~\cite{usenix-enigma2019-android-insider-attack-resistance,url-verified-boot-spec}. In combination with server side validation, this can be used as a form of remote system integrity attestation akin to PCR verification with trusted platform modules (TPMs). 
Integrity of firmware for other CPUs (including, but not limited to, the various radio chipsets, the GPU, touch screen controllers, etc.) is out of scope of AVB at the time of this writing, and is typically handled by OEM-specific boot loaders.

\subsubsection{Verification key hierarchy and updating}
\begin{figure}
	\centering
	\includegraphics[width=0.9\textwidth]{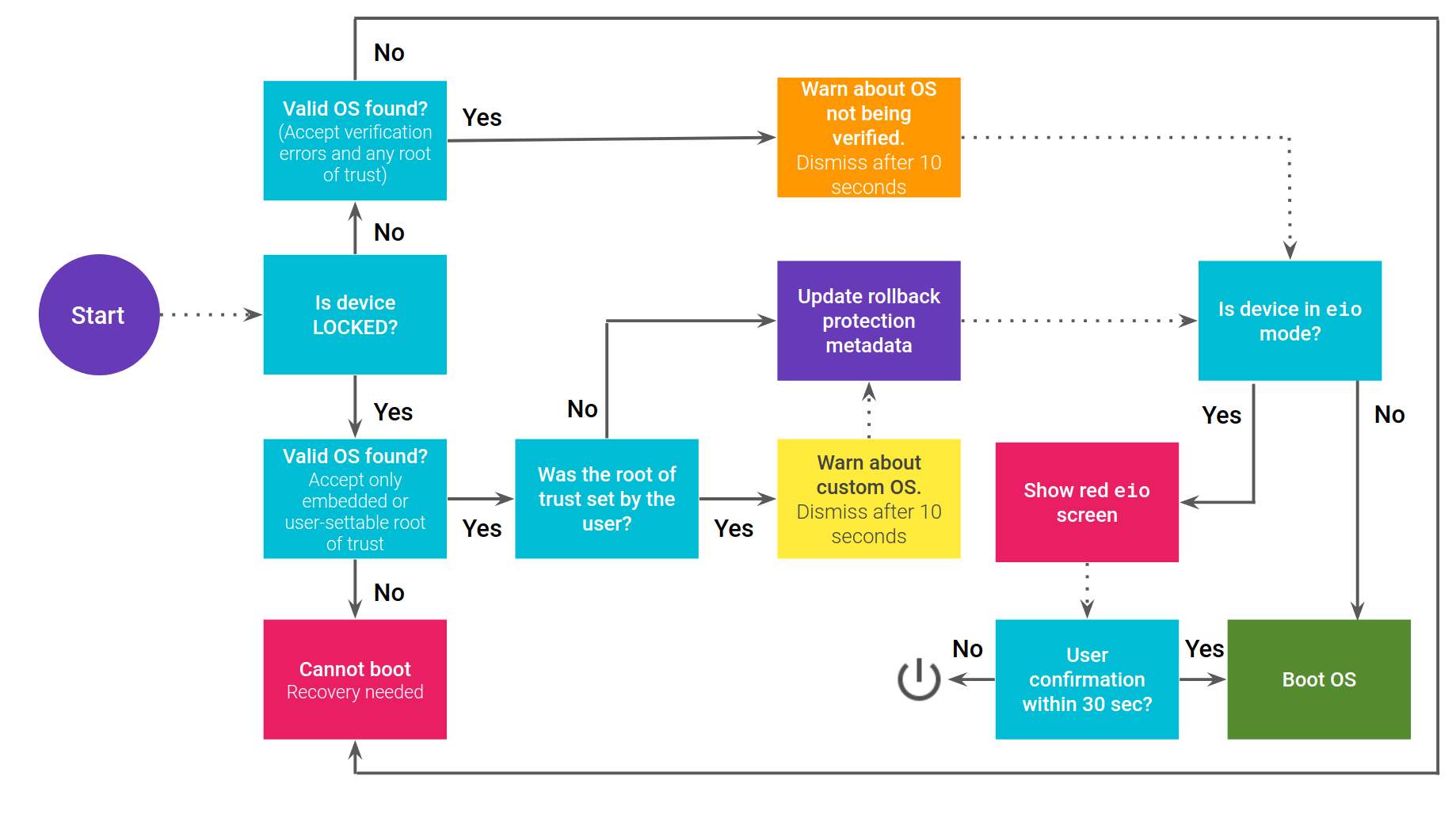}
	\Description[Verified Boot flow and states]{Verified Boot flow and different states}
	\caption{Verified Boot flow and different states: 
		(YELLOW): warning screen for LOCKED devices with custom root of trust set; 
		(ORANGE): warning screen for UNLOCKED devices;
		(RED): warning screen for dm-verity corruption or no valid OS found~\cite{url-verified-boot-flow}.
	}
	\label{fig:verified-boot-flow}
\end{figure}

While the details for early boot stages are highly dependent on the respective chipset hardware and low-level boot loaders, Android devices generally use at least the following keys for verifying system integrity:

\begin{enumerate}
	\item The first (and potentially multiple intermediate) boot loader(s) is/are signed by a key $\mathbf{K_A}$ held by the hardware manufacturer and verified through a public key embedded in the chipset ROM mask. This key cannot be changed.
	
	\item The (final) bootloader responsible for loading the Android Linux kernel is verified through a key $\mathbf{K_B}$ embedded in a previous bootloader. Updating this signing key is chipset specific, but may be possible in the field by updating a previous, intermediate bootloader block. 
	Android~10 strongly recommends that this bootloader use the reference implementation of Android Verified Boot~\cite{url-verified-boot-spec} and VBMeta structs for verifying all read-only (e.g.\ \texttt{system}, \texttt{vendor}, etc.) partitions. 
	
	\item A VBMeta signing key $\mathbf{K_C}$ is either directly embedded in the final bootloader or retrieved from a separate TRH to verify flash partitions before loading the kernel. AVB implementations may also allow a user-defined VBMeta signing key $\mathbf{K_C'}$ to be set (typically in a TEE or TRH) --- in this case, the Verified Boot state will be set to YELLOW to indicate that non-manufacturer keys were use to sign the partitions, but that verification with the user-defined keys has still been performed correctly (see Figure~\ref{fig:verified-boot-flow}). 
	
	Updating this key $\mathbf{K_C}$ used to sign any partitions protected through AVB is supported through the use of chained partitions in the VBMeta struct (resulting in partition-specific signing keys $\mathbf{K_D^i}$ for partition $i$ that are in turn signed by $\mathbf{K_C}$/$\mathbf{K_C'}$), by updating the key used to sign the VBMeta struct itself (through flashing a new version of the final bootloader in an over-the-air update), or -- in the case of user-defined keys -- using direct physical access\footnote{Google Pixel devices support this through \texttt{fastboot flash avb\_custom\_key} as documented online at \url{https://source.android.com/security/verifiedboot/device-state}.}.
	
	\item The digest(s) embedded in VBMeta struct(s) are used by the Android Linux kernel to verify blocks within persistent, read-only partitions on-access using \texttt{dm-verity} (or for small partitions, direct verification before loading them atomically into memory). Inside the \texttt{system} partition, multiple public signing keys are used for different purposes, e.g.\ the platform signing key mentioned in Section~\ref{subsec:permissions} or keys used to verify the download of over-the-air (OTA) update packages before applying them. Those keys can be updated by simply flashing a new \texttt{system} partition.
	
	\item All APKs are individually signed by the respective developer key $\mathbf{K_E^j}$ for APK $j$ (some may be signed by the platform signing key to be granted \texttt{signature} permissions for those components), which in turn are stored on the \texttt{system} or \texttt{data} partition. Integrity of updateable (system or user installed) apps is enforced via APK signing~\cite{url-apk-signing} and is checked by Android's \texttt{PackageManager} during installation and update. Every app is signed and an update can only be installed if the new APK is signed with the same identity or by an identity that was delegated by the original signer. 
	
	For run-time updateable apps, the APK Signature Scheme version~3 was introduced with Android~9.0 to support rotation of these individual signing keys~\cite{url-apk-signing}.
\end{enumerate}

\subsubsection{Integrity and authenticity of system images}
\label{subsec:integrity-gbt}
To mitigate against leaks or insider attacks with resulting control over firmware and system signing keys (questioning if the currently installed factory image is genuine), public tamper-evident records of released versions such as transparency logs can be used to clearly declare which system images were officially released for all users, and therefore to make targeted attacks [T.A8] detectable. E.g., Google Pixel phone firmware has been logged into the Pixel Binary Transparency Log since the release of Pixel~6, and this log can be used to verify that a new firmware update has been created through the official processes and tied to the version booted through verified boot~\cite{androidblogpost-mobile-pixel-binary-transparency}.

\subsection{Patching}
\label{subsec:patching}
Orthogonal to all the previous defense mechanisms, vulnerable code should be fixed to close discovered holes in any of the layers. Regular patching can be seen as another layer of defense. 
However, shipping updated code to the huge and diverse Android ecosystem is a challenge~\cite{10.1145/2808117.2808118} (which is one of the reasons for applying the defense in depth strategy). 

Starting in August 2015, Android has publicly released a monthly security bulletin and patches for security vulnerabilities reported to Google. To address ecosystem diversity, project \emph{Treble}~\cite{10.1145/3358237} launched with Android 8.0, with a goal of reducing the time/cost of updating Android devices~\cite{6676878,androidblogpost-here-comes-treble-2017} and implemented through decoupling of the main system image from hardware-dependent chipset vendor/OEM customization. This modularization introduced a set of security-relevant changes:
\begin{itemize}
	\item The SELinux policy is no longer monolithic, but assembled at boot time from different partitions (currently \texttt{system} and \texttt{vendor}). Updating the policy for platform or hardware components can therefore be done independently through changes within the relevant partition~\cite{aosp-treble-selinux,talk-dcashman-lss-2017}.
	\item Each of the new Hardware Abstraction Layer (HAL) components (mainly native daemons) runs in its own sandbox and is permitted access to only the hardware driver it controls; higher-level system processes accessing this hardware component are now limited to accessing this HAL instead of directly interacting with the hardware driver~\cite{androidblogpost-shut-the-hal-2017}.
\end{itemize}

As part of project Treble, approximately 20 HALs were moved out of the system server, including the HALs for sensors, GPS, fingerprint, Wi-Fi, and more. Previously, a compromise in any of those HALs would allow gaining privileged system permissions, but in Android~8.0, permissions are restricted to the subset needed by the specific HAL. Similarly, HALs for audio, camera, and DRM have been moved out of \texttt{audioserver}, \texttt{cameraserver}, and \texttt{drmserver} respectively.

In 2018, the Android Enterprise Recommended program as well as general agreements with OEMs added the requirement of 90-day guaranteed security updates~\cite{url-android-recommended-security}.

Starting with Android~10, some core system components can be updated independently\footnote{\url{https://source.android.com/docs/core/ota/modular-system}} through the Google Play Store (or through a partner-provided OTA mechanism) as standard APK files or --- if required early in the boot process or involving native system libraries/services --- as an APEX loopback filesystem in turn protected through \texttt{dm-verity}~\cite{androidblogpost-mainline-2019}.

\subsection{Cellular network security}
\label{subec:cellular-network-security}
While Android as a platform and user-focused operating system does not directly depend on the security of any network connection layer (cf.\ Section~\ref{subec:encryption-in-transit}), the majority of Android devices use cellular networks as part of their core functionality. It is therefore important to consider security aspects of this specific network layer as an important use case dependency. There is no robust way for a device to verify the legitimacy of a cellular base station or the integrity of messages it transmits prior to the Authentication and Key Agreement (AKA) handshake~\cite{hussain2019insecure}. This applies to 2G, 3G, 4G, and 5G, and it is a systemic cellular protocol issue, not an OEM or carrier issue. As a result, all mobile devices that support cellular connectivity are susceptible to False Base Station (FBS) attacks~\cite{EFF_gotta_catch_em_all}.

Encrypting data in transit does not protect against all threats derived from an untrusted network. An adversarial cellular network (e.g.\ a FBS) can disable link level encryption at the cellular channel, for example, by silently downgrading the connection to a legacy protocol with weak encryption and no mutual authentication~\cite{practicalattacksLTE}, or by forcing the use of null ciphers. This exposes in the clear circuit-switched Short Message Service (SMS) and voice user communications regardless of TLS encryption of data in transit.
Even when link level encryption is strictly applied to user traffic, e.g.\ Voice over LTE (VoLTE), encryption implementation flaws can allow a passive eavesdropper to decode the encrypted datagrams~\cite{rupprecht2020callmemaybe}.
FBS attacks can also extract or intercept user private identifiers, such as the International Mobile Subscriber Identity (IMSI), regardless of whether all IP-based traffic is end-to-end encrypted.

Android's in-depth security model introduces security features that mitigate [T.N3], as listed in Table~\ref{tab:network-security-improvements}. For example, the option to disable the 2G radio at the modem level~\cite{EFF_victory_2G_toggle} and an option to refuse null-ciphered connections~\cite{androidblogpost-cellular-security}.

Additionally, since 2021 Android has engaged with ecosystem partners to encourage and aid them in hardening their firmwares with modern exploit mitigations ~\cite{androidblogpost-hardening-connectivity-firmware} As a result, starting in Android~13 the platform innovates in leveraging these same compiler-based security features (UBSAN) in connectivity firmware to harden the security of the cellular baseband in some devices (Tables~\ref{tab:application-sandboxing-improvements} and~\ref{tab:system-sandboxing-improvements}).
\section{Special cases}
\label{sec:special-cases}

There are some special cases that require intentional deviations from the abstract security model to balance specific needs of various parties. This section describes some of these but is not intended to be a comprehensive list. One goal of defining the Android security model publicly is to enable researchers to discover potential additional gaps by comparing the implementation in AOSP with the model we describe, and to engage in conversation on those special cases.

\begin{itemize}
	\item \textbf{Listing packages:} The ability for one app to discover what other apps are installed on the device can be considered a potential information leak and violation of user authorization (\hyperref[rule:1]{rule~\circled{1}}). However, app discovery is necessary for some direct app-to-app interaction which is derived from the open ecosystem principle (\hyperref[rule:2]{rule~\circled{2}}). As querying the list of all installed apps is potentially privacy sensitive and has been abused by malware, Android~11 supports more specific app-to-app interaction using platform components and limits general package visibility for apps targeting this API version. While this special case is still supported at the time of this writing, it requires the \texttt{QUERY\_ALL\_PACKAGES} permission and may be limited further in the future.
	
	\item \textbf{VPN apps may monitor/block network traffic for other apps:} This is generally a deviation from the application sandbox model since one app may see and impact traffic from another app (\emph{developer} authorization). VPN apps are granted an exemption because of the value they offer users, such as improved privacy and data usage controls, and because \emph{user} authorization is clear: this is equivalent to the user being responsible for determining the network over which apps will communicate. For applications that use end-to-end encryption, clear-text traffic is not available to the VPN application, which makes it equivalent to the current network operator and keeps the confidentiality of the application sandbox. 
	
	\item \textbf{Backup:} Data from the private app directory is backed up by default. Android~9 added support for end-to-end encryption of backups to the Google cloud by entangling backup session keys with the user lockscreen knowledge factor (LSKF)~\cite{googleblogpost-e2e-backup-2018}. Apps may opt out by setting fields in their manifest.
	
	\item \textbf{Enterprise:} Android allows so-called Device Owner (DO) or Profile Owner (PO) policies to be enforced by a Device Policy Controller (DPC) app. A DO is installed on the primary/main user account, while a PO is installed on a secondary user that acts as a work profile. Work profiles allow separation of personal from enterprise data on a single device and are based on Android multi-user support. This separation is enforced by the same isolation and containment methods that protect apps and users from each other, with a significantly stricter divide between the profiles~\cite{android-enterprise-whitepaper-2023}.
	
	A DPC introduces a fourth party to the authorization model: only if the policy allows an action (e.g.\ within the work profile controlled by a PO) in addition to authorization by all other parties can it be executed. The distinction of personal and work profile is enhanced by the recent support for different user knowledge factors (handled by the lockscreen as explained above in Section~\ref{subsec:lock-screen}), which lead to different encryption keys for FBE. Note that on devices with a work profile managed by PO but no full-device control (i.e.\ no DO), privacy guarantees for the personal profile still need to hold under this security model. Users may choose to turn off the work profile at any time, which causes all associated apps to be stopped and the respective FBE keys to be evicted, ensuring resistance of work profile data against physical attacks in line with [T.P2] even while the personal (main) profile is still actively in use.
	
	\item \textbf{Factory Reset Protection (FRP):} is an exception to not storing any persistent data across factory reset (\hyperref[rule:4]{rule~\circled{4}}), but is a deliberate deviation from this part of the model to mitigate the threat of theft and factory reset ([T.P2][T.P3]).

    \item \textbf{Widevine:} is another exception to \hyperref[rule:4]{rule~\circled{4}}, as its identifier remains stable across factory resets. It allows app developers to detect abuse stemming from device resets, and is scoped to the developer key to prevent cross-app tracking.
	
\end{itemize}

\section{Related Work}
\label{sec:related-work}
Classical operating system security models are primarily concerned with defining access control (read/write/execute or more fine grained) by subjects (but most often single users, groups, or roles) to objects (typically files and other resources controlled by the OS, in combination with permissions sometimes also called protection domains~\cite{Tanenbaum:2014:MOS:2655363}). The most common data structures for efficiently implementing these relations (which, conceptually, are sparse matrices) are Access Control Lists (ACLs)~\cite{312842} and capability lists (e.g.~\cite{capsicum}). One of the first well-known and well-defined models was the Bell-LaPadula multi-level security model~\cite{belllapadula}, which defined properties for assigning permissions and can be considered the abstract basis for Mandatory Access Control and Type Enforcement schemes like SELinux. Consequently, the Android platform security model implicitly builds upon these general models and their principle of least privilege.

One fundamental difference is that, while classical models assume processes started by a user to be a proxy for their actions and therefore executes them directly with user privileges, more contemporary models explicitly acknowledge the threat of malware started by a user and therefore aim to compartmentalize their actions. Many mobile OSes (including Symbian as a historical example) assign permissions to processes (i.e.\ applications) instead of users, and Android uses a comparable approach. A more detailed comparison to other mobile OSes is out of scope for this paper, and we refer to other surveys~\cite{mobile-os-survey1,mobile-os-survey2,7113562} as well as previous analysis of Android security mechanisms and weaknesses exploited by malware~\cite{4768655,Zhang:2013:VUB:2508859.2516689,7446031,li_i_2014,LI201767,7546516,6999911}.

\section{Conclusion}
In this paper, we described the Android platform security model and the complex threat model and ecosystem it needs to operate in. One of the abstract rules is a multi-party authorization model that is different to most standard OS security models in the sense that it implicitly considers applications to have equal veto rights over actions in the same sense that the platform implementation and, obviously, users have. While this may seem restricting from a user point of view, it effectively limits the potential abuse a malicious app can do on data controlled by other apps; by avoiding an all-powerful user account with unfiltered access to all data (as is the default with most current desktop/server OSes), whole classes of threats such as file encrypting ransomware or direct data exfiltration become impractical.

AOSP implements the Android platform security model as well as the general security principles of ``defense in depth'' and ``safe by default''. Different security mechanisms combine as multiple layers of defense, and an important aspect is that even if security relevant bugs exist, they should not necessarily lead to exploits reachable from standard user space code. 
While the current model and its implementation already cover most of the threat model that is currently in scope of Android security and privacy considerations, there are some deliberate special cases to the conceptually simple security model, and there is room for future work:

\begin{itemize}
	\item Keystore already supports API flags/methods to request hardware- or authentication-bound keys. However, apps need to use these methods explicitly to benefit from improvements like Strongbox. Making encryption of app files or directories more transparent by supporting declarative use similar to network security config for TLS connections would make it easier for app developers to securely use these features.
	\item It is common for malware to dynamically load its second stage depending on the respective device it is being installed on, to both try to exploit specific detected vulnerabilities and hide its payload from scanning in the app store. One potential mitigation is to require all executable code to: a) be signed by a key that is trusted by the respective Android instance (e.g.\ with public keys that are pre-shipped in the firmware and/or can be added by end-users) or b) have a special permission to dynamically load/create code during runtime that is not contained in the application bundle itself (the APK file). This could give better control over code integrity, but would still not limit languages or platforms used to create these apps. It is recognized that this mitigation is limited to executable code. Interpreted code or server based configuration would bypass this mitigation.
	\item Advanced attackers may gain access to OEM or vendor code signing keys. Even under such circumstance, it is beneficial to still retain some security and privacy assurances to users. One recent example is the specification and implementation of \emph{Insider Attack Resistance} (IAR) for updateable code in TRH~\cite{androidblogpost-insider-attack-resistance-2018}, and extending similar defenses to higher-level software is desirable~\cite{usenix-enigma2019-android-insider-attack-resistance}. Potential approaches could be reproducible firmware builds or logs of released firmware hashes comparable to e.g.\ Certificate Transparency~\cite{laurie_certificate_2013}.
	\item Hardware level attacks are becoming more popular, and therefore additional (software and hardware) defense against e.g.\ RAM related attacks would add another layer of defense, although, most probably with a trade-off in performance overhead.
\end{itemize}

However, all such future work needs to be done considering its impact on the wider ecosystem and should be kept in line with fundamental Android security rules and principles.

\section*{Acknowledgments}
Previous versions of this paper were co-authored by Nick Kralevich, and we especially thank them for their contributions to Android platform security as well as earlier versions of this text. Additionally, we thank Billy Lau, Joel Galenson, Ivan Lozano, Paul Crowley, Shawn Willden, Jeff Sharkey, Haining Chen, and Xiaowen Xin for input on various parts, and particularly Vishwath Mohan for direct contributions to the Authentication section. We also thank the enormous number of security researchers (\url{https://source.android.com/security/overview/acknowledgements}) who have improved Android over the years and anonymous reviewers who have contributed highly helpful feedback to earlier drafts of this paper.

\bibliographystyle{ACM-Reference-Format}
\bibliography{rene,android,accesscontrolmodels}

\pagebreak
\appendix
\section{Towards a formal notation of Android security model rules}
\label{appendix:formal-rules}
Standard access control models are traditionally based on a matrix notation of $(S,O,A)$ triples with subjects, objects, and a defined set of access permissions $A[s,o]$ (typically read, write, and execute)~\cite{DeCapitanidiVimercati2011-accessmatrix}. While the differences in specific implementations of this conceptual matrix (ACLs vs.\ capabilities) are superfluous for our discussion, the basic notation is becoming limited~\cite{10.1145/605434.605437} and unfortunately not directly applicable to the Android model of multiple stakeholders and combining multiple different types of security controls. 

Within the scope of this first draft of a notation of the Android platform security meta model, we define the involved stakeholders as parties $P \in \boldsymbol{P}$ for a set $\boldsymbol{P}$ with pre-defined classes\footnote{These classes of stakeholders could also be seen as roles in an RBAC notation. However, there is no hierarchical relationship between these stakeholders -- they are intentionally considered to be peers -- and therefore the RBAC notation seems less useful in this case.}:
\begin{itemize}
	\item $P_U$ denotes a user of the system. They may or may not be equivalent to the owner of the hardware (client device such as a smart phone), where hardware ownership is defined as out of scope of the security model at this time. However, users are assumed to own their data. 
	\item $P_D$ denotes the developer of an app, which implicitly includes backend services used by that app. That is, $P_D$ is considered owner of the code that is executed by the app as well as potential owner of data used as part of a service (such as video streaming).
	\item $P_P$ denotes the Android platform or more specifically the set of system components that are neither third-party apps nor representing user data. Examples are cell communication, WiFi or Bluetooth services, standard UI elements, or hardware drivers.
	\item $P_O$ denotes an optional organization that can place additional restrictions on the use of a device, e.g.\ because it is owned by the organization or internal services are accessed through the devices that require these security measures. Examples of relevant organizations are employers or schools.
\end{itemize}

For a specific interaction, e.g. one particular user using one particular app to take a picture and store it on the local filesystem of one particular Android system, the relevant stakeholders will be specific instances of these classes, e.g. $\boldsymbol{P_{photoaction}} = \{P_{U_1}, P_{D_1}, P_{P_1}\}$. This set will usually include 3 or 4 (if an organization policy is in place) specific stakeholders, one from each class. The next interaction may use a different set of specific stakeholders, such as another app, another user (using the same app on the same platform), or continuing the use of the same app by the same user but on a different device. To abstract from those specific use cases, we will use the short form $\forall P$ to refer to all stakeholders of a \emph{current} interaction without loss of generality.

Each stakeholder $P$ has some elements:
\begin{itemize}
	\item $S(P)$ denotes the internal state of this stakeholder in the form of arbitrary data stored and controlled by this party. This can take different forms, e.g.\ files, key/value preferences, streams, etc. Note that the Android platform security model is primarily concerned with internal state stored directly within the respective Android system (temporarily in RAM or permanently on non-volatile storage), and data stored outside the physical system (e.g.\ on cloud services) is considered out of scope. However, internal state of one stakeholder (user account data, an app token, etc.) is often used to directly reference such data, and some rules of the Android platform security model \emph{may} therefore transitively apply to such external data as well.
	\item $C(P, A) \in \{allow, deny\} $ denotes the run-time authorization decision of this party concerning a specific action $A$, which is generally considered to be the context of an authorization query. The specific form of an authorization query varies significantly between stakeholder classes (e.g.\ users will often authorize through a pop-up UI dialog while apps will typically authorize through policies evaluated automatically in their application code) and between actions.
	
	The enforcing agent (e.g.\ platform components acting as security monitors) may cache and re-apply previous authorization decisions without asking again depending on configured policy. For example, user authorization expressed through run-time permissions (which is only one way of users to express consent for a specific class of actions for which such permissions are defined by the platform) can currently result in $C(P_U, A) \in \{allow-always, allow-once, allow-in-foreground, deny-once, deny-always\}$ and stored by the permissions controller for future use. A current $allow$ can therefore result from multiple different responses such as $\{allow-always,allow-once\}$.
\end{itemize}

With these preliminaries, we can more formally specify the access control aspects of Android platform security model rules: 

\paragraph{\hyperref[rule:1]{Rule \circled{1} (multi-party authorization)}} Authorization for executing an action $A$ depends on authorization of all relevant parties.
\begin{equation}
	C(A) = allow \iff \forall P: C(P, A) = allow
\end{equation}
Authorization typically grants (partial) access to the internal state of the stakeholder granting this access. 
\begin{equation}
	C(P, A) = allow \implies S(\forall P) \ni f(S(P), C(A)) 
\end{equation}
where $f(S(P))$ denotes the access control function limiting access to the internal state of $P$ scoped to the context of authorization to the current action $A$. That is, the state accessible to all parties $\forall P$ within the current interaction $A$ includes this additional state of the authorizing party $P$. The type of (partial) access, e.g.\ read or write, depends on the context of an action $A$, but may be explicit in the authorization query (e.g.\ read-only or read-write access permission to external storage). 

Further, authorization of one party may depend on run-time context of another party such as 
\begin{equation*}
	C(P_U,A) = allow-in-foreground \land UI-foreground \in S(P_D) \implies C(P_U,A) = allow
\end{equation*}
where $P_U$ authorization depends on the UI state of $P_D$ within the current interaction. There is currently no complete set of all sub-instances of authorization decisions and their contextual dependencies\footnote{This is especially true for context dependent authorization by apps $P_D$, which can use arbitrarily complex code to implement their own decision process.}, and the potential existence of a set sufficient for expressing all necessary conditions is doubtful.

\medskip

Within the lattice notation of mandatory access control (MAC) policies, this multi-party authorization rule implies trivial lower (no authorization) and upper (all involved stakeholders authorization) bounds. While the BLP model is still an underlying principle of SELinux policies and used for Android sandboxing, it is only a part of the higher-level multi-party authorization: namely authorization expressed by the platform components $P_{P_x}$ is internally derived through through BLP flows. On this level of inter-component permission granting, potential future work could investigate the applicability of the Take-Grant model~\cite{Benantar2005AccessCS} for reasoning about collusion issues. Comparison of the expressive power under a meta model like~\cite{10.1145/605434.605437} or~\cite{10.1145/1542207.1542238} is another potential extension, although a cross-abstraction comparison is at least non-obvious at this point.

\paragraph{\hyperref[rule:2]{Rule \circled{2} (open ecosystem)}} All stakeholder classes $P$ represent unbounded sets, and new specific instances can be created at any time by any party: new users can be created on the system itself, new apps can be published and installed without a central gatekeeping instance, and new platforms (devices) can be created freely as long as they follow these rules (cf.\ \hyperref[rule:3]{rule \circled{3}}). 

\paragraph{\hyperref[rule:4]{Rule~\circled{4} (safe reset)}}
\begin{equation}
	\forall P: S(P) := \emptyset
\end{equation}
For the developer $P_D$, resetting their state is interpreted as uninstalling the app from a specific platform and clearing all app data in the process. For user $P_U$ and platform $P_P$ resetting state implies removing a user account or invoking reset to factory default. The implication is that this also resets all authorization from the resetting party, as authorization is defined as (partial) access to internal state. 

\paragraph{\hyperref[rule:5]{Rule~\circled{5} (applications as principals)}} A developer can have multiple apps, which have distinct internal state. That is, a developer $P_D$ actually manages a set of parties $P_A$ in the form of all apps they sign with their developer key.
\begin{equation}
	P_D \supset \{P_{A_1}, ..., P_{A_n}\}
\end{equation}
Apps within a single developer $P_D$ can explicitly share each other's authorization decisions by requesting to be installed with the same shared UID (which implies signature with the same developer key). One of the key elements of \hyperref[rule:5]{rule \circled{5}}, namely that apps $P_A$ do not implicitly represent users $P_U$ is already enforced through keeping their internal state separate (as defined above in preliminaries).

\paragraph{\hyperref[rule:3]{Rule \circled{3} (compatibility)}} Compatibility is the most difficult to express formally, and we only give an intuitive sketch here. As mentioned in Section~\ref{sec:android-security-model}, the Android platform security model by practical necessity spans multiple layers of abstraction. Compatibility therefore requires a rule on a meta level: all potential instances of user $P_U$, developer $P_D$, app $P_A$, and organization $P_P$ operate under the other rules and can continuously update their state and authorization decisions, while the platform $P_P$ --- specifically as the set of mutually untrusted components enforcing the other rules on different layers --- is pinned to a specific version of AOSP to implement the rules of this model. If $P_P$ fails to fully implement this meta rule, all other rules automatically become invalid. 

That is, \emph{invalidation of any rule leads to invalidation of all others}. Other parties need to learn of an invalid $D_P$ so that they can revoke their own authorization (e.g.\ users re-installing the system image to revert to a known-good state). This directly complements (and effectively enables) \hyperref[rule:2]{Rule \circled{2}} because it allows other parties to trust $P_P$ (which often enforces authorization decisions by these parties). 

On a formal level, enforcement of this rule must necessarily be performed outside the platform security model (hence the elevation to a meta rule) and therefore assumes a trusted third party for platform verification. In the current Android ecosystem, this rule is implemented through a combination of automated test suites (including CTS and VTS, which are available as part of AOSP itself), platform attestation keys provisioned by OEMs and signed by Google for systems verifiably passing those test suites, and APIs to query these attestation results that can be used by the other parties at run-time.

\medskip
Note that this first formalization only captures the access control implication of the model rules. It is subject to future work to evaluate if these rules could be formulated under a meta model like \cite{10.1145/1542207.1542238} and be expressed in tandem with access control models of underlying security controls such as MAC policies in the kernel. However, such an endeavour only seems useful if cross-abstraction validation can then be performed using automated tools.

\end{document}